%% file: manuscript_Cuppen.tex
\documentclass{ar-1col}

\usepackage[comma]{natbib}
\usepackage{url, amssymb, longtable}

\usepackage[version=3]{mhchem}

\usepackage{xr-hyper}
\makeatletter

\newcommand*{\addFileDependency}[1]{
\typeout{(#1)}
\@addtofilelist{#1}
\IfFileExists{#1}{}{\typeout{No file #1.}}
}\makeatother

\newcommand*{\myexternaldocument}[1]{%
\externaldocument{#1}%
\addFileDependency{#1.tex}%
\addFileDependency{#1.aux}%
}
\myexternaldocument{SupplMat_Cuppen}

\jname{Xxxx. Xxx. Xxx. Xxx.}
\jvol{AA}
\jyear{YYYY}
\doi{10.1146/((please add article doi))}

\input{journals}
\begin{document}

\markboth{Cuppen et al.}{Laboratory and Computational Studies of Interstellar Ices}

\title{Laboratory and Computational Studies of Interstellar Ices}

\author{Herma M. Cuppen,$^1$ H. Linnartz,$^2$ and \\ S. Ioppolo$^3$
\affil{$^1$Institute for Molecules and Materials, Radboud University, Nijmegen, The Netherlands, 6525 AC; email: h.cuppen@science.ru.nl}
\affil{$^2$Laboratory for Astrophysics, Leiden Observatory, Leiden University, PO Box 9513, Leiden, the Netherlands, 2300 RA}
\affil{$^3$Center for Interstellar Catalysis, Department of Physics and Astronomy, Aarhus University, Aarhus, Denmark, 8000}}

\begin{abstract}
Ice mantles play a crucial role in shaping the astrochemical inventory of molecules during star and planet formation. 
Small-scale molecular processes have a profound impact on large-scale astronomical evolution. The areas of solid-state laboratory astrophysics and computational chemistry study these processes. We review the laboratory effort on ice spectroscopy;  methodological advances and challenges; and laboratory and computational studies of ice physics and ice chemistry. The latter we put in context with the ice evolution from clouds to disks. Three takeaway messages from this review are

\vspace{1mm}

\begin{minipage}{23.75pc}
\begin{itemize}
\item  Laboratory and computational studies allow interpretation of astronomical ice spectra in terms of identification, ice morphology and, local environmental conditions as well as the formation of the involved chemical compounds. 

\item A detailed understanding of the underlying processes is needed to build reliable astrochemical models to make predictions on the abundances in space.

\item The relative importance of the different ice processes studied in the laboratory and computationally changes along the process of star and planet formation.

\end{itemize}
\end{minipage}
\end{abstract}

\maketitle

\tableofcontents

\section{INTRODUCTION}

\vspace{2mm}
The Universe is littered with the debris of dead and dying stars. This debris includes large quantities of micron and sub-micron-sized dust grains. Large molecular clouds collapse under their own gravitational weight, forming denser and colder cores, in which gas-phase species accrete on dust particles. These dust particles act as extra-terrestrial surfaces on which new molecules can form, creating icy layers that provide the chemical ingredients from which other species can form upon impacting atoms, electrons, and cosmic particles or upon irradiation with UV photons. These interstellar ices are composed of volatile molecules that account for a significant fraction of all the available CNO-group elements in star-formation regions and are presumed to be primary carriers of these elements to protoplanetary disks, where they may become incorporated directly into icy planetesimals or may sublimate and undergo further chemical processing in the gas phase. 

Gas-phase reactions in space are generally not very efficient in creating saturated molecules. This is mainly because the reaction routes leading to the formation of terrestrial-like molecules require a third body to absorb excess energy. The densities, particularly in the interstellar medium (ISM) are simply too low for three-body collisions to effectively occur. Here dust grains offer an alternative: they provide a molecule reservoir on which atoms and molecules can ``accrete, meet, and greet'', i.e., freeze out, diffuse/interact, and react, while the icy surface acts as a third body, stabilizing reactions by absorbing excess energy. Solid-state astrochemistry explores the products, mechanisms, and chemical rates that dominate the involved processes and that result in the formation (or consumption) of relatively simple and abundant molecules, such as \ce{H2O}, \ce{CO2}, CO, \ce{NH3}, \ce{CH4}, and \ce{CH3OH}, as well as larger species, so-called complex organic molecules (COMs, see Box \ref{box:COMs}) that comprise for example of (smaller) alcohols, sugars, and amino acids. Many of these are considered building blocks of life as we know it. Indeed, a substantial fraction of the unambiguously identified molecules in space has a complex organic nature, such as glycolaldehyde (\ce{CH2(OH)CHO}), an important component towards the formation of ribose, dimethyl ether (\ce{CH3OCH3}) and acetamide (\ce{CH3CONH2}), which has been proposed as a precursor of glycine, the smallest amino acid. These COMs are detected through their gas-phase spectra, but are expected to form effectively in the solid state. The involved processes are diverse and clearly different from Earth-based chemistry. Reactions take place at much lower temperatures, which only allows for barrierless reactions or reactions with low barriers that involve tunneling. A large fraction of the involved particles resides in atomic form and the chemical triggers initiating reactions, e.g. through radical formation, are rather exotic, as will be reviewed here. Finally, the involved time scales are very long.

\begin{textbox}[h]\section{COMPLEX ORGANIC MOLECULES}
\label{box:COMs}
Here we define complex organic molecules as highly saturated carbon-containing molecules with at least six atoms, including heteroatoms such as O, N, S, or P. They have hence the general form \ce{C_nH_mX_i} with X=O, N, S, or P and are similar to stable terrestrial molecules that can be bought in a bottle. These COMs are expected to form on grain surfaces, whereas unsaturated species such as carbon chains and cyanopolyenes (\ce{HC_nN}) are largely formed in the gas phase. Our definition of COMs explicitly excludes polycyclic aromatic hydrocarbons (PAHs), even if they contain a heteroatom.
\end{textbox}

Over the past years, substantial progress has been made in understanding the chemical role of inter- and circumstellar ices. This has been achieved through concerted efforts by astronomical observers, astrochemical modelers, laboratory astrophysicists, and computational chemists. The focus of this review is on the laboratory and computational work. There exist two basic approaches to studying ice chemistry in the laboratory: one approach aims at making interstellar ice analogs as realistic as possible with many different components. This approach is relevant from a spectroscopic viewpoint since it gives information on how different components in an ice interact. Chemically, it allows a more representative starting point. It also comes with the difficulty that it is very hard to learn more about individual processes.  The other approach aims at studying a specific process and typically starts with one or two component systems for a series of selected conditions. This is less representative of `real' astronomical ices, but allows one to characterize individual processes in detail that are needed as input for astrochemical models. These models can then extend the impact of these processes to interstellar timescales. 

Writing a review about laboratory and computational studies focusing on interstellar ices is a daunting task. The field is broad and has developed strongly over the last few decades. It ranges from ice spectroscopy to ice chemistry, which can be triggered by atoms, UV, cosmic rays, or heat, and for which different approaches and techniques are needed. This review does not claim to be exhaustive. Rather we aim to give a methodological update highlighting the state-of-the-art, to discuss recent findings for ice processes that are relevant in the star and planet formation process and to look into future challenges and opportunities. This review focuses largely on work performed over the last 10-15 years. Earlier work and work focusing on related topics have been summarized in a number of reviews: \citet{Herbst:2009, vanDishoeck:ChemRev, Boogert:2015, Linnartz:2015, Jorgensen:2020, Oberg:2016}.

This review is organized in the following way. Section~\ref{sec:observations} describes the role of inter- and circumstellar ices from an observational perspective, highlighting the dominant ice processes from cloud to disk. This is followed in Section~\ref{sec:spectroscopy} by a discussion of ice spectra, since this provides the basis for observational identification. Section~\ref{sec:method} discusses several experimental and computational methods. In the subsequent sections the physical properties and processes in ices (Section~\ref{sec:physics}) and surface chemistry (Section~\ref{sec:surfreact}) are discussed. The latter are described following the evolutionary stages of the ice; starting from hydrogenation reactions, and radical-radical reactions to build up water-rich and CO-rich ice layers from the surface to chemistry occurring within the ice layers. The review concludes with a summary and future outlook.

\section{CONSTRAINTS FROM ASTRONOMICAL OBSERVATIONS}
\label{sec:observations}
In an ice matrix molecules cannot rotate, which excludes the use of high-resolution radio and submillimeter observations to identify such species through their rotational spectrum, as in the gas phase. However, molecules can vibrate, to some extent hindered by the surrounding matrix, and hence the direct observation of interstellar ices is nearly fully realized through the detection of characteristic `fingerprint' absorption features in the infrared. The basic concept of such measurements is that one uses the photosphere of a background star as a reference source along a sightline comprising ices in dark clouds or the emission of hot dust or scattered light for ices in protoplanetary disks and protostellar envelopes. Ice bands can be observed by ground-based telescopes, but observations from space are preferred for some species, as many ubiquitous transitions of interstellar features can be obscured by telluric pollution in the Earth's atmosphere, specifically originating from  \ce{H2O} and \ce{CO2}. 
\begin{figure}
    \centering
    \includegraphics[width=\textwidth]{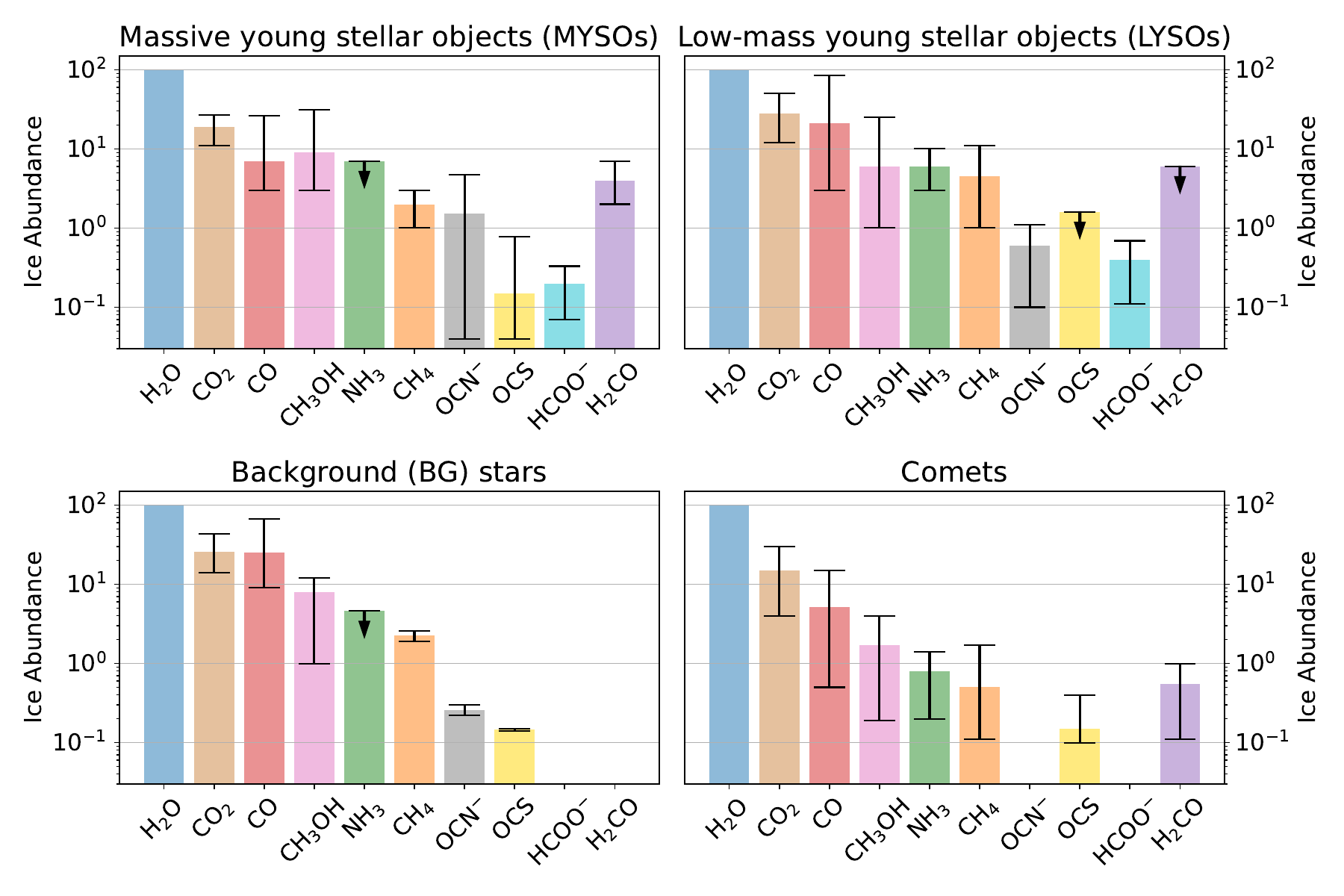}
    \caption{Relative abundance with respect to water (\ce{H2O}) for the nine most abundant ice molecules detected (or likely detected) towards massive young stellar objects, low-mass young stellar objects, background stars, and comets. The black bars indicate the minimum and maximum detected
values and the arrows indicate upper limits. Values are from \citet{Boogert:2015} (most MYSOs and BGs: \ce{CO2}, \ce{CO}, \ce{CH3OH}), \citet{Boogert:2022} (MYSOs: \ce{OCN-}, OCS and LYSOs), \citet{McClure:2023} (BGs: \ce{NH3}, \ce{CH4}, \ce{OCN-}, and OCS), \citet{Mumma:2011} (comets: \ce{CO2}, \ce{NH3}, and \ce{H2CO}), \citet{DiSanti:2008} (comets: \ce{CH3OH}, \ce{CH4}, and CO), and \citet{Saki:2020} (comets: OCS). Updated diagram from  \citet{Rachid:thesis}.
}
    \label{fig:iceabun}
\end{figure}

\begin{figure}
    \centering
    \includegraphics[width=\textwidth]{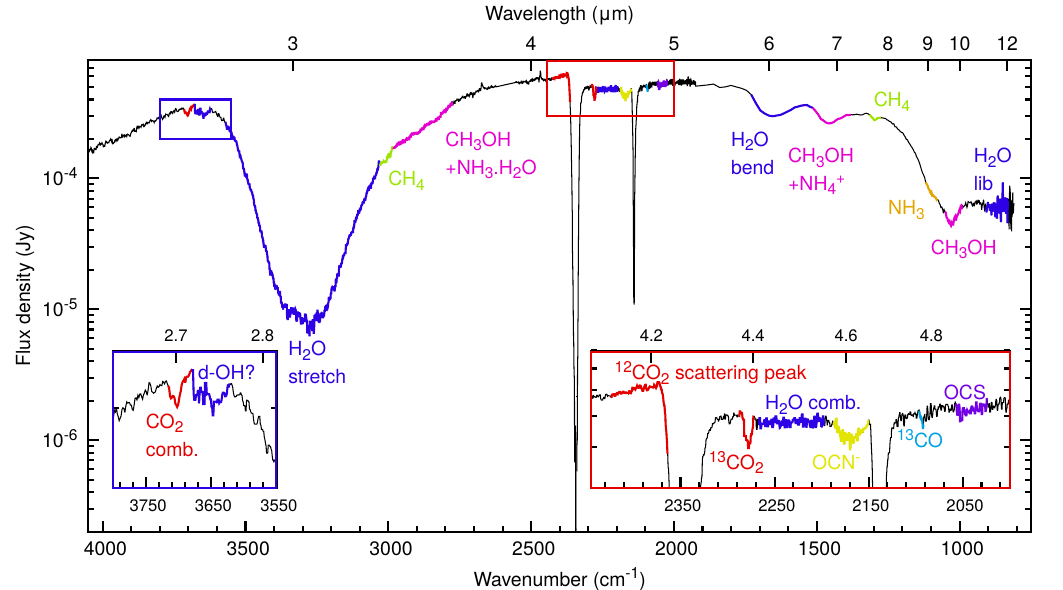}
    \caption{The full IR ice spectrum recorded by JWST towards NIR38 ($A_\text{V} \sim 60$~mag.) in the molecular cloud region Chameleon I. The spectrum shows a large variety of molecular ice signatures that provide detailed information on ice morphology as shown in the insets. The dangling OH bond gives information on the ice structure, the \ce{^{12/13}CO2} and \ce{^{12/13}CO} signals provide information on isotope fractionation, the deviating \ce{^{12}CO$_2$} band profile about grain sizes, and signals in the 6-8~${\mu}$m region cover the range where spectral features are expected of frozen COMs as will be discussed later. Redrawn from: McClure et al., An Ice Age JWST inventory of dense molecular cloud ices, Nature Astronomy, published 2023, Springer Nature.
}
    \label{fig:jwstspectrum}
\end{figure}

Nearly 300 different molecules have been identified in the inter- and circumstellar medium (ISM/CSM) in the gas phase \citep{McGuire:2022}. The number of identified frozen species is much lower, not even making up 5{\%} of the total. \textbf{Figure~\ref{fig:iceabun}} gives the most common ice species, summarizing abundances w.r.t.~water ice towards a number of different sources. Most of the ice observations available are for protostellar envelopes, and are limited to smaller species, like \ce{H2O}, \ce{CO2}, \ce{NH3}, \ce{CH4}, and \ce{CH3OH}. The low number of identified ice constituents may be surprising, at least at first sight, but ice bands are harder to interpret as these are broad, which may result in overlapping features and their peak positions might shift depending on the ice composition. {\textbf{Figure~\ref{fig:jwstspectrum}}} shows a full ice spectrum recorded by the James Webb Space Telescope (JWST) towards NIR38 which is a dense molecular cloud with a $A_\text{V}\sim60$~mag \citep{McClure:2023}. It clearly shows the broad ice features, in particular the \ce{H2O}-stretch band, which has \ce{CH4} and \ce{CH3OH} features on its wings. The tentative dangling OH band (see left inset) possibly originates from the interactions of other ice species with the water-matrix environment, and it offers a good tool to investigate the level of surface structure and mixing of an ice.
JWST, only now, provides the first observational indication of frozen COMs larger than methanol. Their low abundances compared to the ice constituents that dominate ISM ice spectra and the similarity of spectra of different COMs with similar functional groups complicate unique identifications. We will show, how laboratory infrared spectra can be employed to gain insight into morphological parameters of ice in space and how to derive the likely contributions of different species to composite ice spectra.  

\begin{figure}
    \centering
    \includegraphics[width=0.8\textwidth]{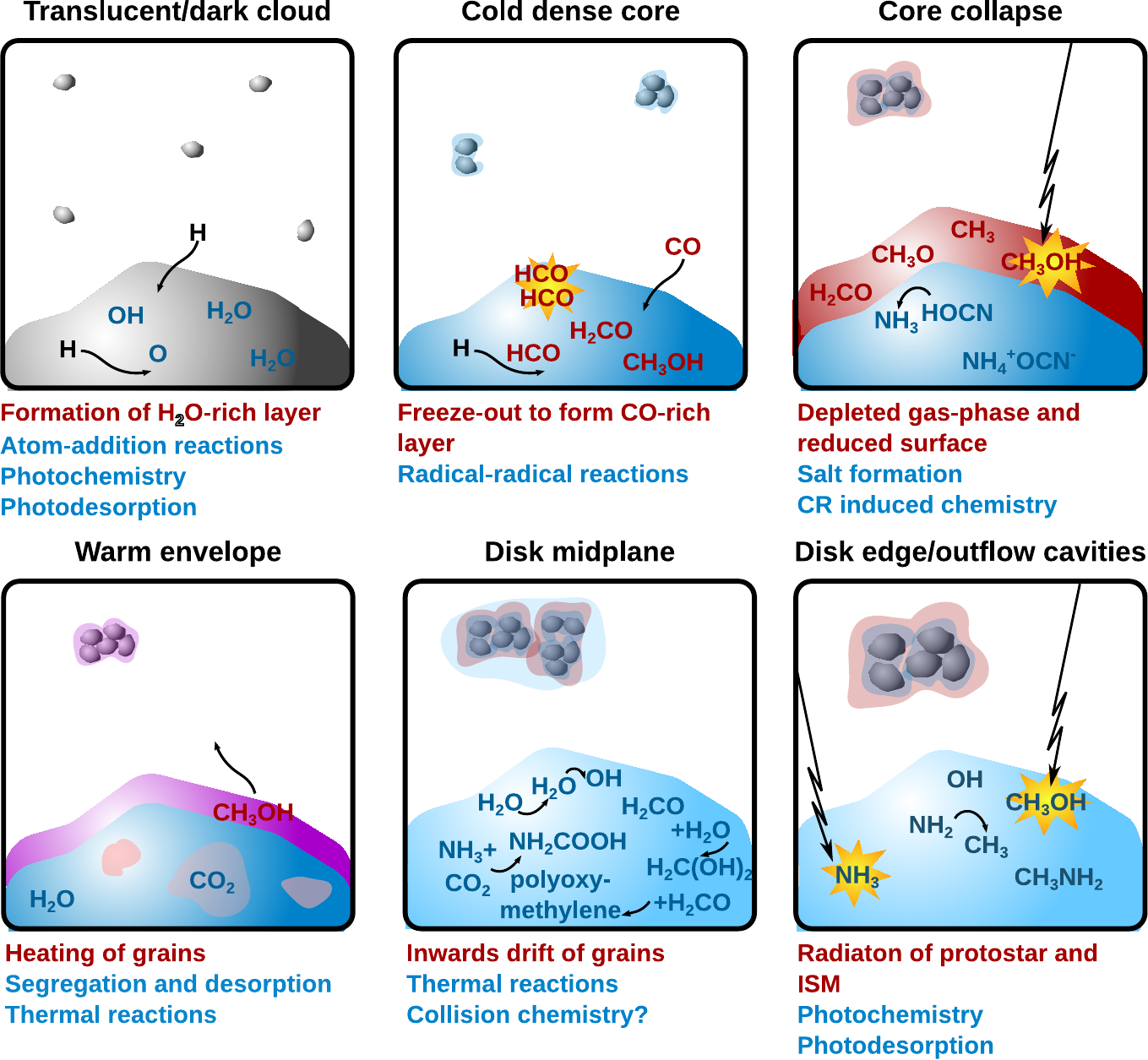}
    \caption{Schematic picture of the evolution of ice mantles across the different evolutionary stages in star and planet formation. Bare grains are in gray, water-rich ice in blue, and CO-rich ice in red. The changing gas-phase composition, grain-surface temperature, and mantle thickness make that different processes dominate during these stages. The picture is based on observational evidence, laboratory experiments, and astrochemical simulations. In the initial molecular cloud (top row) ices build up in layers. In the evolution to a protoplanetary disk, these layers can desorb leaving a residue behind (purple), segregate, or become covered with an additional water ice layer (light blue), depending on the physical conditions. }
    \label{fig:icelayers}
\end{figure}

Comparisons of laboratory spectra of different compositions to observed spectra have revealed that interstellar ice consists of layers, representing different evolutionary stages \citep{Tielens:1991, Boogert:2015}. \textbf{Figure~\ref{fig:icelayers}} summarizes the different layers present in the ice together with the dominant chemical and physical processes at each stage. In principle, all processes occur in all environments but their relative impact changes during the ice evolution. Ices become observable in translucent clouds at roughly $A_\text{V}=1.5$~mag (cloud-to-center) \citep{Whittet:1988}. Here the UV irradiation is attenuated to such a degree that the formation of water ice is faster than its photodissociation and water ice can survive on grain surfaces. In more diffuse areas, ices likely exist up to a few monolayers \citep{Lamberts:2014I}. 

Along the transition from translucent to dense molecular clouds, the gas phase is initially mostly atomic, except for \ce{H2}, and mostly free O-, C-, and N-atoms freeze out onto grains where they become quickly hydrogenated upon H-atom addition leading to a water-rich ice layer that also contains simple ice species such as \ce{NH3} and \ce{CH4} (see Section~\ref{sec:water-rich}, dark blue in \textbf{Figure~\ref{fig:icelayers}}). 
The gas phase becomes slowly depleted of atomic oxygen: part is frozen out onto the grain and converted into \ce{H2O} and another fraction is converted into CO by a series of gas-phase reactions. In this transition phase from O-rich gas to CO-rich gas, \ce{CO2} starts to form on the grain through reactions between \ce{CO} and \ce{OH}, an intermediate in \ce{H2O}-ice formation. \ce{CO2} is indeed found to appear at an $A_\text{V}$ of  3--4~mag, which is between that of \ce{H2O} (3~mag) and \ce{CO} (6~mag). \ce{CO2} ice is further found to reside in mixtures either rich in \ce{H2O} or rich in \ce{CO} \citep{Boogert:2015}. 

In dense molecular cores, all heavy species freeze out onto the grain. Since the main gas-phase species --besides \ce{H2}-- is CO at this point, the ice becomes rich in CO. This catastrophic freeze-out \citep{Pontoppidan:2003a} needs a critical density in quiescent, non-turbulent clouds to afford a sufficient number of collisions between grains and heavy gas-phase species. A layer rich in CO (in red in \textbf{Figure~\ref{fig:icelayers}}) forms on top of the water-rich layer. Reactions relevant to CO-rich ices are discussed in Section~\ref{sec:CO-rich}. Here we only mention that this phase is considered the starting point for COM formation. 

Once this point is reached, most ices have been formed and surface processes become less relevant. The gas phase is depleted and the surface area is reduced due to grain coaggulation, resulting in a much lower influx of new species onto the grains. During this core collapse phase, the temperature is rather low and the chemistry is limited to those reactions with an appreciable rate at low temperatures. Some salts --charge-neutral compounds that consist of cations and anions-- such as \ce{NH4+OCN-} and \ce{NH4+HCOO-} can already form. Cosmic rays -- either directly or indirectly through photon formation -- induce further chemistry. As soon as the protostar starts to form the environment becomes more heterogeneous in terms of physical conditions, which is depicted in the bottom row of  \textbf{Figure~\ref{fig:icelayers}}. 

The protostellar envelope heats up and the ice layers start to segregate into different components. Reactions that are thermally activated also become possible, as well as reactions that require radicals to become mobile within the ice. Once the grains are heated above 100~K in the inner hot corino region, most ice species have thermally desorbed leaving a refractory layer behind (in violet). The gas-phase detections of the desorbing COMs at this stage attest to a rich surface chemistry. The high deuteration fraction that has been observed in IRAS 16293--2422 \citep{Coutens:2016, Jorgensen:2016} as well as the remarkably similar abundance ratios of a number of COMs along different lines of sight \citep[][, see \textbf{Figure~\ref{fig:iceabun}}]{Nazari:2021, vanGelder:2020} suggest that many species share a common chemical history which is in line with solid-state astrochemical processes that already occur during the cold and dark prestellar core phase.

Once a protoplanetary disk has formed, some of the desorbed water condenses out again on the grain surfaces, covering the grains with a thick porous water layer \citep[light blue in \textbf{Figure~\ref{fig:icelayers}},][]{Visser:2009}. Grains move inwards towards the protostar, forming larger and larger grains and pebbles \citep{Lambrechts:2014, Johansen:2021}. Since there is a temperature gradient along the midplane of the disk, thermal reactions, and desorption processes become increasingly important, similar to the processes occurring in the inner warm envelope. Observational evidence for both amorphous and crystalline water ice has been reported toward different protoplanetary disks \citep{Malfait:1998, McClure:2015, Min:2016, Sturm:2023} and protostars \citep{Dartois:1998, Boogert:2008}. At this stage, grain collisions may also trigger reactions, but very little is known about possible mechanisms since this process is hardly studied experimentally or computationally. Finally, at the edge of the disk, in the outflow cavities, and at the edge of the cloud, grains are not shielded and experience radiation from the protostar and the interstellar radiation field, and photo processes leading to chemistry or desorption dominate.

\section{SPECTROSCOPIC PROPERTIES OF ICE}
\label{sec:spectroscopy}
Infrared ice features are clearly different from their gas-phase counterparts; they are typically much broader compared to unperturbed gas-phase spectra, reflecting the inhomogeneity of molecular interactions in the ice matrix. Moreover, peak positions shift as vibrational modes are hindered upon excitation because of interactions with other ice matrix molecules. This is especially true for water ice which has a very broad feature around 3~$\mu$m and is sensitive to the hydrogen bonding environment \citep[e.g.,][]{Noble:2014a, Smit:2017, Noble:2020}. Different molecules have different effects on the shape and peak positions of the ice absorption features. CO mixed with apolar molecules such as \ce{N2}, \ce{O2} and \ce{CO2} has an absorption feature that is blue-shifted with respect to a pure CO ice \citep{Elsila:1997}. Mixing with hydrogen-bonding molecules like \ce{H2O} or \ce{CH3OH} results in red-shifted and broadened features \citep{Sandford:1988b,  Cuppen:2011B}. Different interaction strengths determine the size of these effects which in turn also depend on the type of vibrational interaction. Different vibrational modes -- stretches, bendings, deformations, etc. -- react differently to changes in for example ice composition and temperature; whereas one vibration may gain intensity another may weaken, and whereas one band may broaden another may actually become narrower or even split (i.e., upon crystallization). Such information can be effectively summarized in plots with FWHMs vs. peak position including ice temperatures.

\begin{figure}
\includegraphics[width=0.8\textwidth]{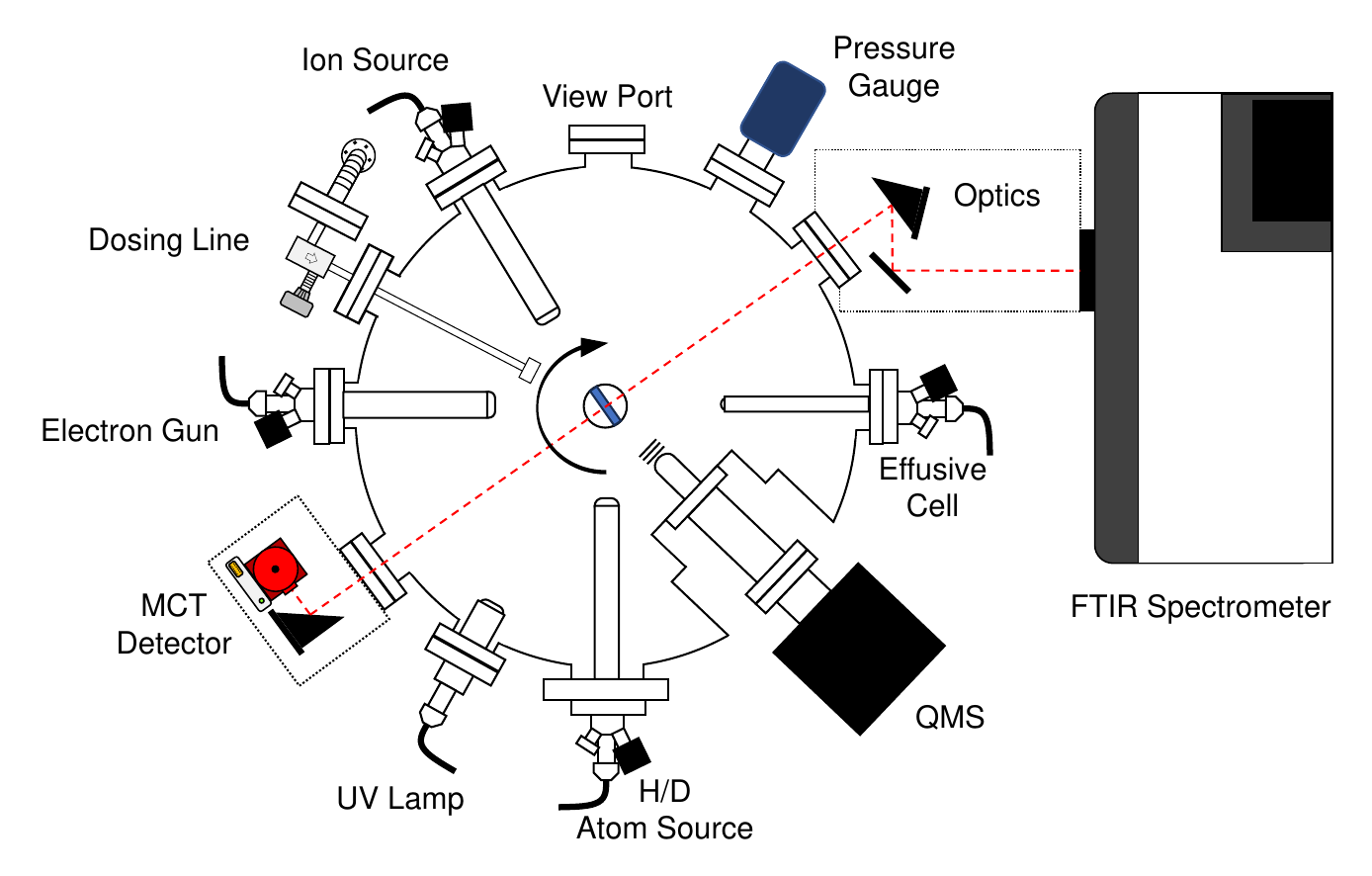}
\caption{Top view of a schematic of a generic ultrahigh vacuum setup optimized to study (non)energetic processing of interstellar ice analogues. The infrared-transparent, rotatable substrate, mounted on top of a cryostat, is  at the center of the chamber. Chemicals are introduced in the chamber through a dosing line and deposited onto the substrate. Effusive cells can be used to deposit solids such as powders. Fourier transform infrared spectroscopy in transmission and mass spectrometry are common analytical tools. A series of processing sources are shown: atom, UV, electron, and ion sources. 
\label{fig:UHV_setup}}
\end{figure}

High-quality vibrational spectra of solid phase molecules in their pure form or as ice mixtures are recorded for temperatures of astrophysical relevance by several laboratory groups, worldwide, and all with the aim to compare the resulting spectra with astronomical data or guide future observations. These experiments are mainly performed in transmission, as transmission spectra can be directly compared to spectra measured towards protostars or background stars. For this, ices are grown under high vacuum (HV, $< 10^{-6}$ mbar) or ultrahigh vacuum (UHV, $< 10^{-9}$ mbar) conditions on a cryogenically cooled and infrared transparent substrate, typically for temperatures as low as 10~K, using closed-cycle He cryostats and applying a method known as background deposition of gases or vapor. A schematic of setup with transmission spectroscopy is shown in \textbf{Figure~\ref{fig:UHV_setup}}. 
The low pressures are needed to reduce the pollution from the freeze-out of residual gas in the vacuum chamber. 
The impact of such pollution effects is generally limited for spectroscopic studies that work with rather thick ices of several thousands of monolayers (MLs).  Thin ice studies, ranging from (sub)monolayer to several tens of MLs require, however, UHV experiments, since accretion rates from background gas at HV conditions are as high as several MLs per minute. The latter is particularly important when studying surface reactions.

The spectral data are obtained by guiding the light of a Fourier Transform Infrared (FTIR) spectrometer through the ice and signals are recorded in direct absorption. The latter needs to be corrected for atmospheric pollutions outside the vacuum chamber which is easily achieved by taking a reference spectrum before ice growth. Typically, spectra are recorded in the 500-4000~cm$^{-1}$ (20-2.5~$\mu$m) mid-infrared range, with extensions into the far infrared (100-500~cm$^{-1}$) \citep{Ioppolo:2014}. The spectral resolution can be as low as 0.1~cm$^{-1}$, which can be considered as high resolution for solid-state spectroscopy. Since this increases recording times over longer wavelength ranges considerably, in most cases ice spectra with 0.5-1~cm$^{-1}$ resolution are obtained. Once deposited at a low temperature, controlled heating allows continuous recording of (changing) spectra for higher temperatures, until the ice or constituents of an ice mixture start to thermally desorb. By repeating this procedure for different ice compositions, a large amount of data is obtained that can be used to fit astronomical observations.  In addition, most spectroscopic ice setups have complementary tools to obtain absolute absorption cross sections, for example, by recording laser interference patterns during ice growth \citep{Bossa:2014} or by incorporating a quartz microbalance \citep{Santoro:2020}. Such information is useful, to derive column densities from astronomical spectra or to predict observing times. 

As mentioned above, spectroscopic ice studies record changes in the peak position, bandwidth, and (relative) band intensity as a function of mixing ratio and temperature. In recent years, the influence of the level of porosity has also been studied \citep{Isokoski:2014}.  
Such spectra have been recorded for many different ices, containing both more abundant (roughly $>3$\%) and less abundant species, such as COMs, and are collected in public ice databases that can be used to interpret and guide astronomical observations. Currently, the largest spectral ice databases available online are the NASA Ice Database by Hudson et al. (https://science.gsfc.nasa.gov/691/comicice/spectra.html) and LIDA, the Leiden Ice Database for Astrochemistry which is accessible through https://icedb.strw.leidenuniv.nl. The latter comprises a steadily increasing number of 1200+ ice spectra with tools that can be used to fit observational data or to simulate spectra. A detailed description and manual is available from  \citet{Rocha:2022}. 
 
Besides pure spectroscopic information, it is also important to have general knowledge about the optical constants of ices, such as the refractive index. This is particularly true for major ice constituents such as amorphous solid water (ASW) and \ce{CO2}. An example is the observed scattering peak shown in the right inset of \textbf{Figure~\ref{fig:jwstspectrum}}. Grain shapes and grain size distribution have a large effect on the overall observed spectrum and the calculation of radiative transfer requires knowledge of the refractive index \citep{Dartois:2022}. It is a complex function consisting of the imaginary index, $k$, that describes the attenuation or absorption of the medium, and the real refractive index, $n$, is the ratio of the velocity of light in the medium with respect to the vacuum speed of light. Both $k$ and $n$ are wavelength-dependent. 
Only very recently, it has become possible to perform laboratory measurements that cover the full wavelength-dependent refractive index values for ASW in the UV--vis range by combining laser and broadband interferometry \citep{Kofman:2019, He:2022B, Stubbing:2020}. The use of stable white light sources, to monitor the interference patterns of individual monochromatic patterns in combination with regular laser-based interferometry of a growing ice, allows us to derive the full $n(\lambda)$ curves, quickly and highly efficiently.
Extensions of such data to the IR have become possible as well, as recently illustrated by \citep{Rocha:2023}. 

We discuss two examples of how laboratory data can be applied for the interpretation of ice observations. The first example links CO and \ce{CH3OH} ice. The CO ice band as found in astronomical observations shows a composite structure, consisting of different contributions that are due to CO interacting with different matrix species. Based on dedicated laboratory measurements, water as an ice partner can be excluded, since even for a large range of parameter settings, a CO:\ce{H2O} ice \cite[see also][]{Fraser:2004, Cuppen:2011B} cannot reproduce the observed spectra. A mixture of \ce{CO}:\ce{CH3OH}, however, does \citep{Cuppen:2011B, Penteado:2015} and hints at a direct chemical connection, fully in line with studies showing that methanol can form efficiently through ongoing hydrogenation starting from CO ice (see Section~\ref{sec:CO-rich}), as will be discussed later. In fact, this CO+H channel is found to be the starting point in the formation of larger and larger COMs, including methyl formate.

\begin{figure}
\includegraphics[width=0.8\textwidth]{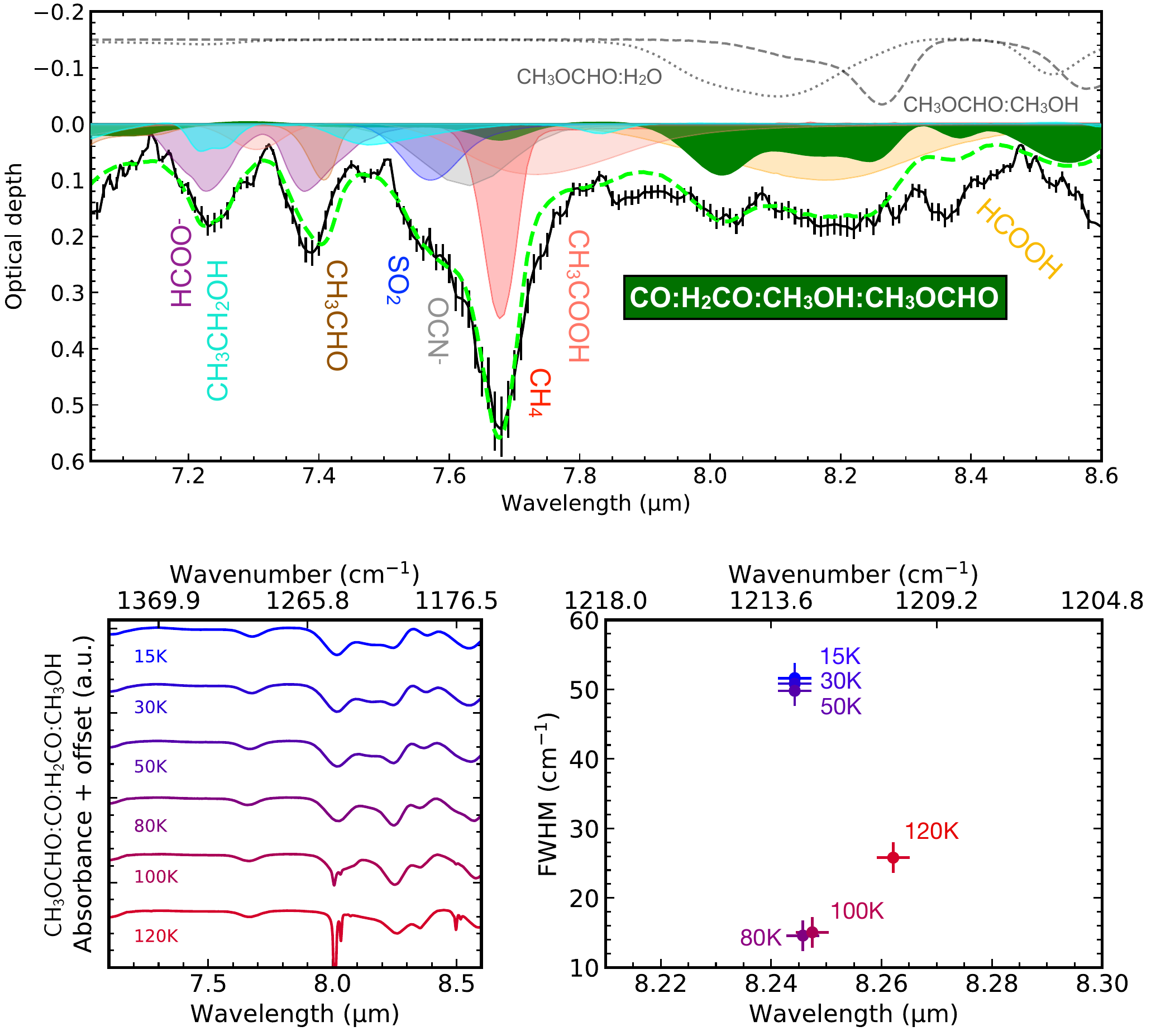}
\caption{A JWST spectrum (black curve) recorded towards IRAS2A zooming in on the COM region (redrawn from \citet{Rocha:sub}). The upper panel shows many individual bands that can be fitted when assuming a mixture of different ice species. The spectrum for methyl formate in a mixture of CO, \ce{H2CO}, and \ce{CH3OH} is shown in green with data presented in \citet{TerwisschaVanScheltinga:2018}. The lower left panel shows the spectral behavior of methyl formate for this mixture as a function of temperature. A scatterplot of the peak position versus FWHM of the C-O stretch feature for different temperatures is given in the lower right panel. Scatterplots of this type can help in selecting the ice components to include in the composite spectrum as shown in the upper panel. It is worth noticing that in the spectrum also clear signals for the anions \ce{HCOO-} and \ce{OCN-} can be seen. For further details see text.}
\label{fig:IR_spectroscopy}
\end{figure}

The second example is shown in \textbf{Figure~\ref{fig:IR_spectroscopy}}. This figure shows how an astronomical spectrum (black curve) can be decomposed into different individual features using laboratory spectra. In the upper panel, a zoom-in is shown of the 7-8.6 $\mu$m region of a JWST spectrum recorded by the JOYS+ (JWST Observations of Young protoStars) consortium towards IRAS2A \citep{Rocha:sub}. The spectrum covers the spectral region where COMs exhibit characteristic vibrational modes. In previous work, clear hints for the presence of frozen COMs (acetaldehyde and ethanol) were reported, both in data from the Spitzer telescope \citep{TerwisschaVanScheltinga:2018} and more recently, in JWST observations towards a Class 0 protostar \citep{Yang:2022} in this region of the spectrum. 
The top panel of \textbf{Figure~\ref{fig:IR_spectroscopy}} shows in light green an overall fit to the spectrum, which is composed of different spectra of 
individual molecules that again may be shifted or broadened with respect to their pure spectra because of matrix interactions or because of different temperatures. Special care has to be taken not to overfit such a spectrum and to decrease ambiguity to an absolute minimum. The fit routine used to obtain the green fit \citep{Rocha:sub} is based on the ENIIGMA code \citep{Rocha:2021} that is also directly accessible through the Leiden Ice Database for Astrochemistry. Besides COMs, spectral signatures are included from \ce{CH4}, \ce{HCOOH}, \ce{SO2}, \ce{OCN-}, and \ce{HCOO-}.
The focus in the figure is on methyl formate (\ce{CH3OCHO}).
In the upper panel, a methyl-formate spectrum is indicated in dark green for an ice mixture containing CO, \ce{H2CO}, and CH$_3$OH. In the left lower panel, laboratory spectra are shown of methyl formate in this mixture for different temperatures. For higher temperatures narrowing and even splittings can be seen. Details are available from \citep{TerwisschaVanScheltinga:2018}. For one specific band, the C-O stretch mode around 1210~cm$^{-1}$ (8.25~$\mu$m) the results are summarized in the lower right panel showing FWHM vs.~peak position for different temperatures. 
For other methyl formate bands (not overlapping with modes of abundant ice species) such as the C=O stretch (1720~cm$^{-1}$), CH$_3$ rock (1165~cm$^{-1}$), O-CH$_3$ stretch (910~cm$^{-1}$), and OCO deformation mode (768~cm$^{-1}$), similar data have been recorded. Scatterplots of the FWHM vs.~peak position for all these transitions and in different ice environments can then aid in the selection of the most likely contributions to the composited spectrum to reproduce the astronomical spectra in terms of ice composition and environmental conditions. 
If multiple bands of a specific ice species are available, these also should show up in the right intensity ratios (unless full spectral overlaps prohibit). If only a single band can be found, this band has to `perfectly' fit the laboratory data. In the case of methyl formate the data show that mixtures with only \ce{CH3OH} and \ce{H2O} are not consistent with the observations (light gray curves in upper panel).
The important conclusion from this JOYS+ work is that apart from methanol, acetaldehyde, ethanol, and methyl formate can be identified beyond doubt as frozen COMs. This makes sense since their chemical solid-state networks are connected. Besides these three COMs a careful claim is made that also acetic acid (\ce{CH3COOH}) features can be identified. It is very likely, that also many other COMs reside in these ices, as is discussed later in this review. For now, their detection is still beyond what is possible. 

\section{METHODOLOGY AND CHALLENGES}
\label{sec:method}
Many of the processes on surfaces are thermally activated with a rate constant 
\begin{equation}
\label{eq:LH_th_barrier}
k = \nu\exp \left(-\frac{E_a}{kT}\right),
\end{equation}
where $E_a$ is the activation barrier,  $T$ the temperature, and $\nu$ the pre-exponential factor. With quantum chemical calculations, the barrier can be determined and both experiments and simulations can be applied to obtain the rate constant $k$. For a rate constant to be relevant or measurable, it has to be sufficiently high for the process to occur on the relevant time scale. For a first-order process, the half-time is given by $\ln(2)/k$ and this half-time can be of the order of thousands of years for the process to be relevant in interstellar space. To be measurable in the lab it needs to be in the hour range and for calculations in the nanosecond range. Thermally activated processes can still be studied by simply increasing the temperature. Desorption of water ice, for instance, occurs around 160~K in the laboratory and around 100~K in space because of these timescale differences \citet{Viti:2004}. 
Below we first look into a generic experimental approach to study ice processing (extending on the spectroscopic work described in section 3) and then we focus on the computational details. 

\subsection{Laboratory perspective}
\label{sec:method_lab}
A typical ice experiment follows a sequence of logical steps: ice formation, characterization, triggering the (non)energetic process(es) of interest, data recording, and interpretation of the results \citep{Linnartz:2015}. The ice is typically prepared from stable and commercially available species (gases, liquids, and solids) by introducing them into the vacuum chamber in gaseous form (see  \textbf{Figure~\ref{fig:UHV_setup}}). The species then freeze out onto the cold substrate and an ice is formed as described in Section~\ref{sec:spectroscopy}. Species can be pre-deposited or co-deposited, to form layered or homogeneously mixed ices. Typical laboratory fluxes are much higher than in space, but temperatures and vacuum conditions as well as ice morphologies are representative for interstellar conditions.  When studying reactions or processes involving radical species, there is the extra challenge of introducing reactive species to the ice under cryogenic and UHV conditions. Reactive species can be created within the ice by energetic processing \citep{Herczku:2021}. Here `energetic' refers to events that deposit more than a few meV of kinetic or photon energy into the ice upon impact.  

UV irradiation and energetic processing lead to dissociation and radical formation in the bulk of the ice, at ice depths that depend on the energy of the projectiles (UV photons, electron, ions) \citep{Oberg:2009A}. Radicals can also be send to the ice directly by radical beams, which is more representative of dark interstellar cloud conditions. The impacting radicals are often `hot' and need to be cooled to ensure that a surface reaction is not caused by excess thermal energy. Cooling, however, also decreases the number of radical species, as collisional cooling is applied \citep{Watanabe:2004, Ioppolo:2013a}. 
In the earlier days, predeposition was the standard where first an ice was grown that was exposed to radicals in a second step \citep{Watanabe:2004,Fuchs:2009}.  Currently, co-deposition experiments in which, for instance, radical and the ice species land on the surface simultaneously, have become more common \citep{Cuppen:2010B}. This approach has three main advantages: \emph{(i)} intermediate species can be ``frozen in'' depending on the radical/ice ratio and \emph{(ii)} the surface is constantly refreshed allowing more material to be converted. These advantages enhance the abundance levels of species that are hard to detect, either because they are transient species, are inefficiently formed, or have low band strengths. Moreover, \emph{(iii)} this method reflects better ice growth in space. 

Once the reactive species are on or in the ice, new species can form, and detection techniques are required to record the consumption of reactants or the formation of new species. The two most common tools are temperature programmed desorption quadrupole mass spectrometry (TPD-QMS) and transmission (or reflection absorption) infrared spectroscopy (TIRS/RAIRS). Quadrupole mass spectrometry can identify species by their mass, upon desorption from the ice surface. Different species desorb at different temperatures. This allows to perform a temperature-programmed desorption experiment: mass spectra are recorded while linearly increasing the temperature. This technique can detect fractions of monolayers and is hence very sensitive, but only upon desorption of the ice. Typical surface experiments are performed at low temperatures of 10~K, whereas desorption of some products occurs only above $\>100$~K. One has to disentangle the low-temperature chemistry from the possible thermal chemistry during the TPD. TIRS and RAIRS are less sensitive than QMS, but can detect species directly in the ice and hence also at low temperatures. TIRS is mostly used to investigate the bulk of the ice that is usually 10s to 1000s nm thick as already mentioned in Section~\ref{sec:spectroscopy}, whereas RAIRS is mostly employed to characterize ices from the submonolayer layer regime to several tens of monolayers. Often infrared spectroscopic and mass spectrometric detection techniques are used simultaneously as they are complementary. The use of isotopic precursor species adds another tool to draw unambiguous conclusions, as this affects both QMS and RAIRS data. For a review on other analytical techniques currently used in laboratory ice astrophysics we refer to \citet{Allodi:2013}.

The introduction of reactive species in or on interstellar ice analogs opens the opportunity for a network of reactions to occur. Experimental abundances result from all reactions in this network in competition with diffusion or with each other. It is hard to disentangle these different processes. Models can help in this sense, as well as probing different reaction conditions. Choosing different H/CO ratios in the study of CO hydrogenation for instance, and using \ce{H2CO} or \ce{CH3OH} as starting material instead of CO probes different parts of the reaction network and can help in disentangling the complexity of the network \citep{Cuppen:2010B}. To obtain reaction rate constants, one does not only need to disentangle the different processes, but also have a knowledge on the UV or particle fluxes which is far from trivial to obtain \citep[see e.g.,][]{Ioppolo:2013a, Ligterink:2015}.

A barrier, typically of a few 1000~K,  renders a reaction impossible at the cold interstellar temperatures, as highlighted by Eq.~\ref{eq:LH_th_barrier}. For reactions involving H atoms, tunneling through the reaction barrier greatly enhances the probability of reaction. Performing an experiment twice using either H or D is a convenient way to detect the difference in tunneling efficiency. If the reaction products scale with the effective mass of the reaction, this is usually interpreted as proof of tunneling and this effect is called the Kinetic Isotope Effect (KIE). For both reactions, isotope substitution can be done on either of the two reactants, leading to four different combinations. An extra complication is that the mass does not only affect the reaction rate. The sticking probability of D atoms to ice is higher and their binding energy is stronger. This leads to a higher D atom coverage on the surface with respect to H atom coverage. The experimentally determined KIEs are a result of all these effects combined and are hence expected to be less pronounced than can be expected based on the pure reaction rate. Computationally, one can study the KIE of the isolated reaction.

\subsection{Computational perspective}
\label{sec:method_comp}
Quantum chemical calculations play  an important role in the interpretation of laboratory results or in determining input values which are critical for astrochemical models and we therefore  explain in more detail the different types of calculations and what can be taken from those. A wide variety of different methods, levels of theory, and descriptions of the systems are used, which makes it hard to compare different studies and value the different results, especially for non-experts. In Box~\ref{box:compmethods} a brief definition of the different methods is given.

\begin{textbox}[h]\section{COMPUTATIONAL CHEMISTRY METHODS}
\label{box:compmethods}
\vspace{-3ex}
\textbf{\textcolor{fignumcolor}{Different levels of theory}}\\

\begin{description}
    \item[Force fields] use analytical expressions to describe the interaction between atoms and molecules. They are parameterized to reproduce experiments or quantum chemical calculations.
    \item[Density Functional Theory] uses a quantum mechanical description of the energy. Interactions between electrons are not treated explicitly, but in an approximated way using an effective electron density, described by the functional.
    \item[Coupled cluster] is another quantum-chemistry method which uses multi-electron wavefunctions to account for electron correlation. This method is expensive in terms of computational time, but can accurately describe weak van der Waals interactions. 
\end{description}
    \textbf{\textcolor{fignumcolor}{Different simulation methods}}
\begin{description}    
    \item[Static calculations] only explore the energy landscape. Energies, barriers, and frequencies can be obtained. 
    \item[Molecular dynamics calculations] solve the classical equations of motion and include temperature effects and energy flows. Rates of fast processes and frequencies can be determined.
    \item[Kinetic Monte Carlo] simulates diffusing, desorption, and reaction processes. It ignores local vibrations to reach larger timescales.
\end{description}

\end{textbox}

Because of computational constraints, the ideal quantum chemistry method does not exist and one has to make compromises at several stages. 
One way to compromise is by reducing the system size. Although an interstellar particle is small, its size is still much larger than typical systems in computational chemistry calculations which range from roughly 20~{\AA} in diameter for DFT calculations to 100~{\AA} for force field calculations. These systems can be cluster models or periodic models, where periodic boundary conditions make the surface artificially infinitely large. Accurately considering the periodic interactions in the latter case, adds a significant computational demand. Cluster models are hence an interesting cheaper alternative, but they cannot capture the full complexity in binding environments, as discussed in Section~\ref{sec:desorption}. 

Another obvious compromise is in the level of theory. Usually, Density Functional Theory (DFT) is used, but the functionals in DFT come in many different flavors with different accuracies and very different computational demands. Functionals are typically categorized according to ``Jacob's Ladder'', from the \emph{down-to-Earth} Hartree-Fock Theory to the \emph{Heaven} of chemical accuracy. Although this is an oversimplification and sometimes ``lower rung'' functionals accidentally lead to better numerical results for some specific systems, it is a good starting point in interpreting DFT results. The lower rungs on the ladder will result in good geometries, but higher rungs are needed for suitable barriers, binding energies, or frequency calculations. DFT is not so good at describing dispersion interactions; this is the attractive part of the van-der-Waals interaction and is the dominant interaction in apolar ices such CO and \ce{CO2}. Dispersion corrections exist, and again they come with different accuracies and computational costs. Coupled-cluster calculations at the ``gold standard level'' (CBS/CCSD(T)) are very good at obtaining the dispersion interaction but come at a very high computational price which scales badly with system size. They were traditionally only used for gas-phase calculations, but recent developments now make these calculations affordable for solid-state systems as well. Force fields are computationally very affordable and scale nicely with system sizes. They can, in principle, be adjusted to reproduce the most important features of high-level calculations. 
There is a trend towards multiscale calculations, often called QM/MM or QM/QM calculations, where a periodic system is calculated at a lower level of theory (MM(force field) or low-level QM), but corrected locally at a higher level of theory (QM) for the more interesting part, for instance, the binding site. Some examples are discussed in Section~\ref{sec:desorption}. 

A third area to compromise is in terms of dynamics, one could replace dynamic calculations with static calculations that determine energy barriers and infer rates from those. Static calculations do not include temperature effects or dissipation of energy and are less accurate in that sense. However, since they are computationally less demanding, higher-level calculations for the energies and forces compared to dynamics calculations can be applied. Dynamics simulations are typically done using classical trajectories of the nuclei and at each time step forces on all atoms need to be calculated. 

Finally, quantum effects can be included or not. We have already discussed quantum mechanical calculations for the energy and forces, but other effects such as quantized energy transfer and tunneling can be important as well. The dynamics of the atoms in the ice is typically described using classical dynamics which ignore zero-point energy and the quantized nature of energy levels. Tunneling through a reaction barrier is generally recognized as important since it can greatly increase the probability of a reaction. This occurs through delocalization of the transition state. To accurately describe tunneling, a full quantum-mechanical calculation is preferred and there are numerous methods available to calculate tunneling rates \citep[as summarized by][]{Kaestner:2014}. In astrochemical models often approximations are used such as the Wentzel-Kramers-Brillouin approximation 
for tunneling through a rectangular barrier, which only uses the barrier height as an input parameter. However, the exact shape of the barrier and the effective mass of the reaction coordinate, which is not always directly apparent, also play a role. As a consequence, the resulting tunneling rate constant as obtained from the approximated expression can be orders of magnitude off with respect to rate constants that have been calculated in a proper quantum-mechanical treatment \citep{Goumans:2010a}.

\section{PHYSICAL PROPERTIES AND PROCESSES IN ICES}
\label{sec:physics}
The following subsections describe the physical properties, governing the  processes introduced in \textbf{Figure~\ref{fig:icelayers}}. The reactive processes in this figure are described in Section~\ref{sec:surfreact}.

\subsection{Adsorption}
\label{sec:adsorption}
Ice surface chemistry starts with the adsorption of species onto the surface. Three important factors in this respect are the cross-section of the grain, the sticking orientation of the molecule, and the sticking fraction of atoms and molecules to the grain. For low gas and grain temperatures, the sticking fraction of most species, except light species like \ce{H} and \ce{H2}, is close to unity, but when an incoming particle needs to lose a large amount of kinetic energy because of its high incoming velocity the sticking fraction can be much lower. \citet{Chaabouni:2012} show experimentally that the sticking coefficient of physisorbed \ce{H2} increases linearly with the number of deuterium molecules already adsorbed on the surface. Here the deuterium molecules help absorb some of the excess kinetic energy. Computationally, sticking fractions of molecules have been determined by Molecular Dynamics \citep[e.g.,][]{Buch:1991, Takahashi:1999, Al-Halabi:2004, Veeraghattam:2014}. 

The cross section for accretion is typically assumed to be the geometrical cross-section of the grain. However, long-range interactions between the grain and a gas-phase molecule can enhance the accretion cross-section. \citet{Cazaux:2022} invoked the interaction between negatively charged grains and \ce{S+} to explain the depletion of sulfur in denser regions. The attraction is described in a free molecular form where the grains are assumed to be spherical with the electric potential to be symmetrical distributed on the grain. Measurements of collision rate constants for charged aerosol particles with ions \citep{Pfeifer:2023, Gopalakrishnan:2012} show that in the limit of low densities and low excess charge, this is indeed a reasonable description. The model by \citet{Cazaux:2022} then assumes that \ce{S+} is immediately neutralized on the grain, but computations on \ce{HCO+} reaching a negatively charged grain show that neutralization is not so trivial as it might seem and can also lead to dissociation into H and CO \citep{Rimola:2021}.

Sticking of heavier molecules, like \ce{H2O} and CO, results in the build-up of an ice. The final structure of the ice depends on the sticking orientation and the possibility for the molecules to rearrange upon adsorption. Water, for instance, with its strong hydrogen-bonded interactions shows ballistic deposition and the porosity of the final structure depends on the incoming angle \citep{Kimmel:2001a, Cazaux:2015A}. This is especially important to understand the difference between ices in the laboratory formed by deposition and ices in space that are typically formed through reactions. The latter are most likely compact \citep{Oba:2009, Accolla:2013}.
Molecules with a small electric dipole align their dipoles when deposited at low temperatures, resulting in an electric field over several layers \citep{Balog:2009, Plekan:2017}. CO is one of the molecules to show this so-called spontelectric behavior \citep{Lasne:2015}. This impacts the structure of the resulting ice, which could have implications for the surface chemistry. Moreover, the electrostatic force of the surfaces acting on gas phase particles can also alter the collision cross-section.

\subsection{Surface and bulk diffusion }
\label{sec:diffusion}
Once on the grain, reactants need to meet to react. One way to do this is through the diffusion of one or both reactants. 
Obtaining diffusion barriers experimentally is challenging. Typically, diffusion rates cannot be measured directly, but have to be inferred from another process. This is often a diffusion-limited reaction where the diffusing species and an immobile reaction partner are deposited. The diffusion rate can then be inferred from the appearance of the reaction product or the disappearances of the reaction partners. This is under the assumption that the dose of both reactants is accurately known since these are required to determine the diffusion length and that the diffusion-limited reaction is indeed the dominant process.
This approach is limited to reactive species and requires a very sensitive detection technique to keep the dose of the reactants low and hence the diffusion length long. 

Most information is available for the diffusion of H atoms on ASW \citep{Watanabe:2010, Hama:2012, Kuwahata:2015} and CO ice \citep{Kimura:2018}. The results show that H diffusion on ASW is highly coverage-dependent. When H atoms initially land on the surface they can move rapidly using shallow binding sites ($E_\text{diff} \leq 210 \pm 20$~K) and middle binding sites ($E_\text{diff} = 260 \pm 10$~K) until they find deep binding sites where they will remain trapped ($E_\text{diff} = 350$~K). Long-range diffusion simulations \citep{Asgeirsson:2017} and binding and diffusion barrier calculations \citep{Senevirathne:2017} confirm this picture. If deep adsorption sites are present, long-distance diffusion which is required to scan a large part of the surface is determined by the escape rate from the deep wells. Here tunneling does not play a significant role. However, if the deep sites are blocked by other adsorbates --in agreement with simulations of CO diffusion on ASW \citep{Karssemeijer:2014I}--, tunneling becomes important for the diffusion between sites.

Another method for obtaining diffusion rates that is more suited for stable species is to deposit a layer of porous ASW on top of an ice consisting of the species of interest.
The species in the bottom layer diffuse into the ASW layer over the surface of pore walls and if the temperature is high enough the species will eventually desorb once it reaches the top of the layer. Diffusion can be inferred by
tracing either the diffusion into the ASW layer or the desorption spectroscopically.  Surface diffusion rates for CO molecules have been studied this way 
\citep{Mispelaer:2013, Karssemeijer:2014I, Lauck:2015, Cooke:2018}. This type of experiment is likely a good proxy for the desorption of CO in the midplane of protoplanetary disks where it is covered with a porous water ice layer \citep{Simons:prep}. For centimeter-size grains, CO desorption starts indeed to become rate limiting compared to the inwards drift of the grains in the disk.

Finally, recent experimental studies use direct microscopic measurements to measure diffusion. Here the mean diffusion distance is measured from the distance between islands that form on the surface using transmission electron microscopy for different temperatures. This can be related to the diffusion coefficient. \citet{Kouchi:2020} measured the activation energy for surface diffusion of CO and \ce{CO2} in this way. For CO diffusion, they found a barrier of $350 \pm 50$~K which is in agreement with the value by \citet{Karssemeijer:2014I}, but significantly higher than the value of $158 \pm 12$~K by \citet{Lauck:2015} and lower than $490 \pm 12$~K by \citet{He:2018}.

Computationally, there have been quite a number of efforts in calculating binding energies (see Section~\ref{sec:desorption}). Studies on diffusion barriers are much less common. This is because they are substantially harder to calculate. On a surface, a range of diffusion barriers exists that can be crossed classically \citep{Batista:2001} or by tunneling \citep{Senevirathne:2017}. Long-time scale calculations need to be performed to determine which diffusion processes are rate-limiting to reach long-scale diffusion. These simulations are rather time-consuming and are typically done by kinetic Monte Carlo \citep{Karssemeijer:2012, Karssemeijer:2014III, Pedersen:2015, Karssemeijer:2014I, Asgeirsson:2017}, since Molecular Dynamics simulations do not reach sufficient time scales and can only be applied for a small number of systems \citep{Ghesquiere:2015}. Advanced simulation methods like metadynamics can help overcome this issue \citep{Zaverkin:2021}.

Since diffusion rates are hard to determine, both experimentally and computationally, compared to binding energies, often a fixed ratio between the binding energy and the diffusion barrier is applied to determine diffusion barriers from binding energies. 
There is no fundamental physical argument why such a universal ratio should exist and it is used solely due to the lack of data.
\citet{Karssemeijer:2014III} determined this ratio for \ce{CO}, \ce{CO2}, and \ce{H2O} on different water surfaces and found  a more or less constant ratio between $0.3-0.4$. A recent study by \citet{Furuya:2022A} showed no universal value. In this work, diffusion barriers were obtained for a number of species using electron microscopy, and binding energies were taken from the literature. The binding energies came however from a diverse set of resources with very different uncertainties and the large range in diffusion/desorption ratios between the different species could be due to the inhomogeneity in binding energy data. Using more homogeneous binding energies \citep{Minissale:2022, Ligterink:2023} shows, however, again deviations from a universal value, especially for larger species.

\subsection{Thermal desorption, binding energies, and trapping}
\label{sec:desorption}
The binding energy of a species determines at which temperature a large fraction is released from the grain and enters the gas phase. Here we only briefly touch on this topic, since an extensive review fully dedicated to binding energies was recently published \citep{Minissale:2022}, along with a list of recommended values \citep{Ligterink:2023}. 
The thermal desorption rate depends on the binding energy of the species to the surface, $E_{\text{bind,}A}$,  
\begin{equation}
 k_{\text{sublimation,}A} = \nu \exp\left(-\frac{E_{\text{bind,}A}}{kT}\right),
 \label{eq:evap}
\end{equation}
where $\nu$ is the pre-exponential factor. Both can be determined experimentally using Temperature Programmed Desorption (TPD) experiments. In these experiments, the substrate is dosed with a known quantity of the species of interest at low temperatures. After deposition, the substrate temperature is linearly increased and the number of desorbing species as a function of time --and hence temperature-- is recorded. The recorded TPD spectrum is then fitted to the Polanyi-Wigner equation
\begin{equation}
\frac{\text{d}}{\text{d} T}n_\text{g}(A) =  n_\text{s}(A)^o \frac{\nu_\text{exp}\exp(- E_{\text{bind,}A} / k T)}{\beta}
\label{Polanyi}
\end{equation}
where $o$ is the order of the desorption process and $\beta$ is the heating rate to obtain the experimental prefactor $\nu_\text{exp}$ and the binding energy. These two quantities are correlated and reliable information can only be obtained if TPD spectra are measured for different initial doses and/or different heating rates to avoid a degeneracy in the fit to obtain both $E_{\text{bind,}A}$ and $\nu_\text{exp}$. As can be seen from Eq.~\ref{Polanyi}, the unit of  $\nu_\text{exp}$ depends on the order of the desorption, whereas  Eq.~\ref{eq:evap} only treats first-order desorption leading to a pre-exponential factor in s$^{-1}$. In astrochemistry often either a constant pre-factor of $10^{12}$~s$^{-1}$ is used or an expression that depends on the binding energy and effectively decreases with the size of the molecules. Experimental studies that carefully extracted both quantities show that the pre-exponential factor increases with the size of the molecule and is much more in line with a prefactor derived from transition state theory (TST) \citep{Tait:2005, Minissale:2022} than the two treatments used in astrochemistry. Experimentally obtained binding energies using these -- more correct-- prefactors are lower than those that are obtained when simply assuming a pre-factor of $10^{12}$~s$^{-1}$. On experimental time scales, this will not result in a sufficient difference for the temperature at which desorption occurs, but for interstellar timescales or when determining snowlines, this can make a substantial difference. \citet{Minissale:2022} show for instance the snowline of \ce{CO2} moves from roughly 40 to 30~AU with a $M_\odot=1$ disk using the updated desorption data.
\cite{Ligterink:2023} determined the binding energies based on literature TPD data for many different species using the corrected prefactor. The new data contain many larger COMs, that have not been previously reported in astrochemical literature. 
With the large amount of binding energies available, machine-learning strategies that predict binding energies for missing species have become within reach \citep{Villadsen:2022}.

In UHV setups, desorbing molecules are effectively pumped away upon desorption and re-adsorption is negligible at the relevant timescale of the TPD experiments. In space, this is not necessarily the case and the sublimation temperature can depend on the local pressure since it affects the adsorption/desorption balance. {\textbf{Table~\ref{tab:Tsub}} gives the sublimation temperature for a list of selected species under dense molecular cloud conditions ($n_\text{H}= 10^5$~cm$^{-3}$) and in the midplane of a protoplanetary disk with a density of  $n_\text{H}= 10^{10}$~cm$^{-3}$. The sublimation temperature is calculated by equating the adsorption and desorption rates, assuming a sticking probability of unity, a site density of $10^{15}$ sites cm$^{-2}$, the same relative composition of the gas and the ice mantle with respect to \ce{H2O}, and assuming $n_\text{gas}(\ce{H2O}) = 10^{-4} n_\text{H}$. In all cases, the sublimation temperature in the midplane is higher than in a dense cloud and the difference can be tens of degrees. The binding energies in \textbf{Table~\ref{tab:Tsub}} are on a ASW surface. Since CO generally resides in a CO-rich layer, the sublimation temperature from a pure CO ice is given as well. The sublimation temperature of \ce{CH3OH} from ASW is higher than that of ASW itself (see \ce{H2O} in \textbf{Table~\ref{tab:Tsub}}) because of its high prefactor. This effectively means that \ce{CH3OH} desorbs with \ce{H2O} since its multilayer desorption temperature is below that of water \citep{Smith:2014}.

\begin{table}[h]
\tabcolsep7.5pt
\caption{Recommended prefactors and binding energies for selected species to ASW and their sublimation temperatures in two different environments.}
\label{tab:Tsub}
\begin{center}
\begin{tabular}{@{}l|c|c|c|c@{}}
\hline
Species & $\nu$$^{\rm a}$ & $E_\text{bind}$$^{\rm a}$& $T_\text{sub,cloud}$$^{\rm b,c}$&$T_\text{sub,midplane}$$^{\rm b,d}$\\
  & {(}s$^{-1}$) & {(}K) & {(}K) &{(}K)\\
\hline
\ce{CH3OH} &$3.18 \times 10^{17} $ & 6621 & 103.9 & 127.0\\
\ce{H2O} & $4.96 \times 10^{15}$ & 5705 & 96.1 & 119.6\\
\ce{NH3} & $1.94 \times 10^{15}$ & 5362& 91.8& 114.6\\
\ce{CO2} & $6.81 \times10^{16}$ &3196 &51.0 & 62.6\\
CO from ASW& $9.14\times 10^{14} $& 1390 & 23.7& 29.6\\
CO from CO & $9.14\times 10^{14} $& 1070$^{\rm e}$ & 18.2 & 22.7\\
\ce{CH4} & $5.43 \times 10^{13}$ &  1232 & 22.2 & 28.1\\
\ce{N2} & $4.51\times 10^{14}$ & 1074 & 18.5 & 23.2\\
\hline
\end{tabular}
\end{center}
\begin{tabnote}
$^{\rm a}$ taken from Table~2 in \citet{Minissale:2022}; $^{\rm b}$ temperature at which desorption and adsorption balance using the assumptions $T_\text{grain}=T_\text{gas}$, $S=1$, and $f_\text{grain}=10^4f_\text{gas}$; $^{\rm c}$ $n_\text{H} = 10^5$~cm$^{-3}$; $^{\rm c}$ $n_\text{H} = 10^{10}$~cm$^{-3}$; $^{\rm e}$ data from \citet{Fayolle:2016} using Redhead approach
\end{tabnote}
\end{table}

Many TPD experiments are of pure ices and the obtained binding energies are binding energies of the species to itself. Another option is to perform TPD experiments of thin layers within the submonolayer regime and then the binding energy of the underlying substrate can be obtained. As discussed earlier, interstellar ices contain a mixture of different species. The introduction of more species in the ice immediately increases the complexity of the desorption process. It is generally not possible to derive binding energies for the desorption of these mixed ices in the same way as for pure ices.  The binding energy of individual species will vary depending on their surrounding material, and the dominant ice-mantle species can prevent other species from desorbing.  \citet{Collings:2003b} showed, for instance, that a large fraction of CO  can become trapped in water ice and is released at higher temperatures. CO needs to diffuse out of the water matrix in order to desorb and hence the desorption can become diffusion-limited (see Section~\ref{sec:diffusion}). This process competes with pore collapse \citep{Bossa:2012, Isokoski:2014, Bossa:2015, Mitterdorfer:2014, Hill:2016} that traps species in the water matrix. Segregation \citep{Oberg:2009A, Noble:2015, Nguyen:2018} impacts the desorption process as well; for instance by facilitating the liberation of species from the water matrix. Crystallization of the water matrix can lead to a molecular volcano effect that pushes out trapped species \citep{Smith:2011}. All these processes are in competition and shape the ice matrix from which the desorption occurs and affect the desorption process. Due to the vastly different timescales in the laboratory compared to space conditions, the competition between these processes might play out very differently \citep{Oberg:2009A}. It is hence essential to understand the microphysics of these competing processes. Dedicated ice trapping experiments can help with this \citep{Bergner:2022, Fayolle:2011A, Fayolle:2016}.

\begin{figure}
\includegraphics[width=0.8\textwidth]{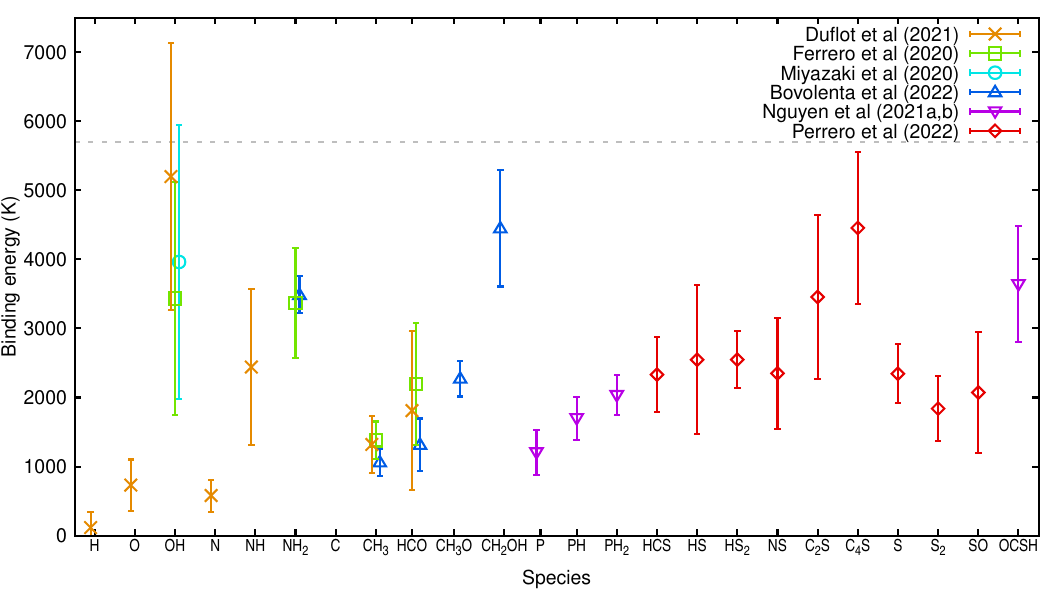}
\caption{Calculated binding energy for a selected number of radical species to ASW. The bars indicate the distribution in binding energies as explained in the text. The dashed line indicates the recommended binding energy for ASW.}
\label{fig:Ebind}
\end{figure}

It is difficult to experimentally obtain binding energies for radical species due to their high reactivity (and correspondingly short lifetime).   
However, several studies report the experimental determination of the binding energy of atomic oxygen on a range of surfaces \citep[][]{Dulieu:2013, He:2014, He:2015A}, showing that for some species, direct measurements are possible.  Computationally, it is easier to obtain binding energies for radical species, although not trivial due to their open-shell character. \textbf{Figure~\ref{fig:Ebind}} shows the binding energies or radical species on an ASW surface taken from several computational studies. Table \ref{SItab:binding} gives the numerical values as well as the prefactor following the treatment of \citet{Tait:2005} and results for binding to a crystalline water ice.

All studies agree that there is a large distribution of binding energies on ASW.
\citet{Bovolenta:2022} determined binding energies for 225 different configurations per adsorbing species distributed among 12-15 ASW clusters consisting of 22 water molecules. They then fitted their obtained distributions of binding energies by one or two Gaussians; the symbols and bars in the plot represent the $\mu$ and $\sigma$ values of the dominant Gaussian. The other studies used periodic models with a QM/QM or QM/MM approach and  significantly fewer binding configurations were considered because of the computational costs (4 to 10). These types of studies with a high level of theory and with a full periodic model were not deemed possible only a few years ago, but are now more routinely performed as is apparent from the reported values in Table~\ref{SItab:binding}.
The average binding energy and standard deviation are used for all cases except \citet{Ferrero:2020}, where $\sigma$ is approximated by the half difference between the minimum and maximum binding energy for their four binding sites.
A direct comparison between the studies is therefore not straightforward since the 4 to 10 configurations used might not be representative of the total distribution of binding sites and only capture a tail of the total distribution. 
Considering this argument, the similarity of the results of \citet{Ferrero:2020} and \citet{Duflot:2021} in \textbf{Figure~\ref{fig:Ebind}}, who use very similar approaches, is promising, at least  for \ce{CH3} and HCO. The results for \ce{OH} show a larger discrepancy between the different studies. This can be because \ce{OH} is a hydrogen bonding species and the binding sites are sampled differently in the three studies. \citet{Miyazaki:2020} and \citet{Ferrero:2020} manually selected different binding sites and subsequently found the local minimum energy configuration, whereas  \citet{Duflot:2021} used molecular dynamics trajectories where the adsorbates were allowed to move around the surface a little to find a stable binding site. This could have led to more stable binding sites. These molecular dynamics simulations are more representative of how adsorbates find their binding sites in space. 

The method by \citet{Bovolenta:2022} searches systematically for different binding sites and should hence cover the full range of binding sites. Their results appear to give lower $\sigma$ values, however. This is most likely because these calculations are performed on cluster models with only 22 water molecules and miss the possibility of sampling local binding pockets due to the high curvature of the clusters. Moreover, the calculations are performed on a lower level of theory to reduce the computational costs and allow for the large number of binding configurations to be considered and the larger number of adsorbates in their binding energy database. Several other studies using much smaller clusters of four to seven molecules exist in the literature, but these are even less likely to reflect realistic binding sites on a water surface.  We have hence not added these values to Table~\ref{SItab:binding}.

Astrochemical models typically only consider a single binding energy value, which does not reflect the results in \textbf{Figure~\ref{fig:Ebind}}. Properly including a distribution of binding sites is however not so straightforward, since diffusion can occur between sites and this will become site-dependent as well. Attempts to formulate rate equations to treat this in a consistent way also show that it is highly dependent on how these different binding sites are distributed on the surface \citep{Cuppen:2011A}.

\subsection{Photodesorption}
\label{sec:photodesorption}
For several decades, different non-thermal desorption mechanisms have been suggested and studied \citep[e.g.,][]{Leger:1985, Hartquist:1990}. This work was triggered by the observation of residual gas-phase CO in dense dark clouds, where full CO depletion was expected because of the low temperature and high densities. The initial focus was on UV photodesorption and desorption caused by cosmic-rays. Later also other (non-dissociative) desorption mechanisms, such as reactive desorption or thermal co-desorption, attracted attention as a way to transfer ice species to the gas phase. 

Photodesorption is by far the best-studied non-thermal desorption mechanism. Upon absorption of a photon, a (sub)surface molecule is excited with sufficient energy to overcome its binding energy or to transfer this energy to another surface molecule that subsequently desorbs. Most studies have focused on the impact of vacuum UV photons, typically in the 120-170 nm region. Such photons are usually generated by broadband, microwave discharge hydrogen-flow lamps (MDHL) that mimic the interstellar radiation field (ISRF), and have fluxes of the order of $10^{14}$ photons cm$^{-2}$s$^{-1}$. The exact UV flux and emission spectrum depend on the lamp settings \citep{Ligterink:2015}. 
The best-studied photodesorption process is for (pure) CO ice, not only because of its astronomical relevance, but also because CO does not ionize or fragment upon VUV excitation, which makes it an ideal molecule to investigate in the laboratory and theoretically. 

A typical photodesorption experiment registers the decrease in TIRS/RAIRS CO ice signal with fluence (i.e. time $\times$ VUV flux). Alternatively, also a mass-spectrometric signal increase can be used, or a combination of the two approaches. As the signal can be transferred into a molecule number, this allows the conversion of the signal change into a photodesorption rate \citep{Oberg:2007B}. The very first theoretical studies predicted CO photodesorption rates of the order of $10^{-6}$ molecules per photon (i.e., roughly one million photons are needed to desorb one CO molecule), but in a series of experimental studies, it was found that this value is substantially higher and lies in the $10^{-2}$ to $10^{-3}$ range. These values are in the same range as recent theoretical predictions \citep{vanHemert:2015}. The precise values obtained by different groups, however, are not fully consistent, stretching from 0.85 to $5 \times 10^{-4}$ molecules/photon, see e.g. \citep{Chen:2014I, Oberg:2007B, Paardekooper:2016}. 
These differences can be largely explained by different experimental conditions, that can be critical, specifically the MDHL settings \citep{Paardekooper:2016}.
The use of monochromatic light at synchrotron facilities such as SOLEIL-DESIRS and ASTRID2 alleviates the aforementioned lamp-to-lamp difference in the emission spectrum. Wavelength-dependent studies showed that the desorption rate followed one-to-one the rovibronic absorption profile of the A-X electronic band system \citep{Fayolle:2011B}. It was concluded that the photodesorption process is governed by a process known as DIET: desorption induced upon electronic excitation. 

Many more photodesorption studies have been performed over the last 1.5 decades, reporting rates for \ce{H2O}, \ce{CO2}, \ce{NH3}, \ce{CH4}, \ce{O2}, \ce{H2CO}, \ce{CH3OH} and other ices, as summarized in Table~\ref{tab:PD-rates}. 
For \ce{H2O} as for many of the other ice species mentioned above, the desorption rates are harder to determine, since VUV irradiation also induces photodissociation; a depleting signal of a specific ice species with irradiation may have a different origin than photodesorption. Instead, the molecule dissocciates and the resulting fragments may photodesorb. The fragments may also chemically react, forming new species that may photodesorb as well. A signal decrease for a precursor species, therefore, should not be taken one-to-one as due to photodesorption. Computationally, it is much more straightforward to disentangle the different desorption channels which has be done in a series of Molecular Dynamics studies  which study isotope effects, structure effects, and the dependency on ice temperature
 \citep[e.g.,][]{Koning:2013, Arasa:2015, Arasa:2010}. An excited molecule was found to be able to `kick out' a neighboring molecule from the ice \citep{Andersson:2008}, which was later confirmed experimentally \citep{Yabushita:2009}. 

Experimentally, \citet{Oberg:2009C} decoupled photodesorption and photochemistry of methanol by fitting its decreasing RAIRS signal by a zeroth-order and first-order process, respectively. Unfortunately, the photodesorption of pure methanol is rather small ($<$ 10$^{-4}$ molecules/photon) \citep[see also][]{Cruz-Diaz:2016}, and most of the desorption takes place through photofragments as nicely shown in a wavelength dependent study by \citep{Bertin:2016}. This means that VUV photodesorption does not offer an efficient pathway to transfer methanol in one piece from ice to gas phase. The same is expected to hold for COMs, i.e., these species will desorb but fragmented. The pure photodesorption rate also can be disentangled from the photodissociation process by comparing VUV irradiated ices with and without an Ar coating on top, combined with laser ablation, as demonstrated recently for \ce{CO} and \ce{H2O} ice \citep{Paardekooper:2016, Bulak:2023}.
Table~\ref{tab:PD-rates} lists the available laboratory-based photodesorption rates (in the vacuum UV and IR) from the literature. It should be noted though, that these values have to be used with care. The experimental settings in different experiments - such as temperature, ice thickness and used irradiation source - may vary substantial. Moreover, flux calibrations and the aforementioned distangling of photodesorption and photodissociation turned out to be hard. This all adds to rather large uncertainties in photodesorption rates of typically 25\% or higher. Finally, one has to realize that these photodesorption studies are largely derived for pure ices. The studies on mixed ices are rather limited.

\input{table_PD}

In mixed ices not only direct but also indirect desorption mechanisms may be at work, such as co-desorption, where the photo-excitation of one molecule causes another species to desorb. 
For mixed ices of CO:\ce{N2} it was shown that this is highly effective: upon excitation of CO, \ce{N2} photodesorbs and upon excitation of \ce{N2}, CO photodesorbs \citep{Bertin:2013}. However, such an indirect photodesorption process does not work for CO mixed with methanol, at least not upon VUV excitation.  The slightly higher methanol mass compared to \ce{N2} does not explain this observation. It is more likely that the higher binding energy prohibits an efficient co-desorption. For these reasons, co-desorption is excluded for the moment as an effective non-thermal desorption mechanism to explain for example COM gas phase abundances.

The question remains how non-thermal desorption processes can transfer ice species, including COMs, into the gas phase in a non-dissociative way. The answer may be found in different wavelength regimes. X-ray photodesorption experiments of \ce{H2O} showed efficient desorption of neutral \ce{H2O} whereas ionized \ce{H2O} was only a minor component \citep{Dupuy:2018}. Similarly, X-ray photodesorption of CO leads predominantly to neutral CO, with minor ionic channels such as \ce{O-}, \ce{C-}, and large molecular cations \citep{Dupuy:2021}. Even large molecules like acetonitrile (\ce{CH3CN}) and methanol can efficiently desorb intact in this way with desorption rates around $10^{-2}-10^{-4}$ molecules/photon for desorption from pure or CO-rich ices \citep{Basalgete:2021II}. Desorption is much less efficient from \ce{H2O}-rich ices. The fraction of intact desorbing molecules appears to be dependent on the X-ray flux since the destruction process is more efficient for higher fluxes \citep{Basalgete:2021II, Ciaravella:2020}.

A promising wavelength regime for intact desorption is the infrared. IR irradiation studies on  water ice showed indeed that water can desorb upon resonant IR irradiation. This is especially efficient for crystalline water ice, whereas irradiation of ASW and \ce{CO2} leads predominantly to restructuring \citep{Noble:2020, Cuppen:2022, Ioppolo:2022}.  In experiments focusing on the IR photodesorption of CO, \ce{CH3OH}, and mixed CO:\ce{CH3OH} ices \citep{Santos:2023A} it was found that the IR photodesorption rates are many orders smaller compared to VUV irradiation -- of the order of $10^{-8}$ molecules/photon -- but it was also concluded that the astronomical impact can be substantial, possibly  even higher than in the VUV, as in the ISM the flux of IR photons may exceed that of VUV photons by several orders of magnitude. This is particularly interesting because, upon IR excitation, the chance that a molecule dissociates is also much smaller. 

Finally, the impact of cosmic rays can result in desorption. Desorption of mantle species as a consequence of grain heating has been considered relevant from an astrochemical perspective for already quite some time \citep{Hartquist:1990,Hasegawa:1993A, Herbst:2006}, but desorption can also occur through radiolysis, spot heating, or sputtering\citep{Dartois:2021, Dartois:2018, Wakelam:2021}. In very recent work, a literatur e overview is provided of the cosmic rays sputtering yields for molecules known to reside in interstellar ices \citep{Dartois:2023}. Experimental insights into the microphysics of sputtering and molecule destruction were reported for CO ice bombarded by cosmic rays \citet{Ivlev:2023}. Given the high energies involved upon impact, this mechanism may be closer to ice sputtering than ice desorption. To some extent this is comparable to mantle destruction through shocks or jets around forming stars \citep{Lee:2019, Zeng:2020}, a mechanism that is hard to study experimentally.

\subsection{Reactive or chemical desorption}
\label{sec:reactivedesorption}
The binding energy of COMs is rather high and it hence came as a surprise that molecules such as dimethyl ether (\ce{CH3OCH3}), methanol (\ce{CH3OH}), and methyl formate (\ce{HCOOCH3}) were detected in the gas phase of dark molecular clouds, i.e., at low temperatures where these species should be frozen out and with very low UV fluxes \citep{Bacmann:2012, Bacmann:2016, Jimenez-Serra:2016, Cernicharo:2012}.  This triggered studies on different non-thermal desorption mechanisms such reactive or chemical desorption. The idea behind this is that reaction products of a surface reaction might be  excited and that this excitation energy can lead to the desorption of the reaction products, even at low temperatures. Indeed, chemical models that account for this mechanism in some way have observed an increase in the COMs in the gas phase \citep{Vasyunin:2017, Wakelam:2017, Fredon:2021A, Drozdovskaya:2014, Cuppen:2017, Cazaux:2015B}. The initial treatment of this mechanism was purely theoretical where Rice--Ramsperger--Kassel--Marcus theory was used to arrive at an expression for the desorption probability \citep{Garrod:2006b, Garrod:2007}. 

\citet{Dulieu:2013} undertook the first experimental verification through sequential \ce{O2} hydrogenation experiments, leading to \ce{H2O2} and \ce{H2O}, on 
an amorphous silicate or a graphite surface. They found substantial desorption of the formed \ce{H2O} molecules in the submonolayer regime, which is likely due to the lack of interactions with surrounding molecules, but a significant drop in the desorption probability in multilayer regime, \emph{i.e.}, on an ice surface \citep{Minissale:2014}.  For other reaction systems (e.g., CO~+~H, \ce{H2CO}~+~H), they determined relatively low reactive desorption rates in the (close-to) sub-monolayer regime  \citep[$\lesssim 10$\%,][]{Minissale:2016}.  Using a collisional model they obtained an expression that was in reasonable agreement with their experimental data for rigid surfaces like graphite, but it showed poorer agreement for flexible surfaces like water ice \citep{Minissale:2016}. Since then several other groups have confirmed the efficiency of reactive desorption to release heavy surface species to the gas phase \citep{Oba:2018, Oba:2019, Chuang:2018, Nguyen:2021A, Nguyen:2020, Santos:2023C}.

Experimental verification of expressions for chemical desorption is not straightforward for several reasons: experiments measure typically desorption probabilities ``per reactive species'' whereas models require probabilities ``per reactive event'' \citep{Oba:2018} and systematic variation of parameters is challenging since a change of reactive system leads to change in many relevant parameters. In all cases, models are required to interpret the results and it is crucial to fully understand the system to obtain reliable results from such a model as was already stressed in Section~\ref{sec:diffusion}. The recent work by \citet{Santos:2023C} exemplified this. Here loss of HS and \ce{H2S} ice was found to be not only by reactive desorption, as was assumed in earlier studies, but also by \ce{H2S2} formation. The two competing loss routes could be quantified. 

In a computational approach, it is easier to control the different parameters involved in chemical desorption as is done in classical Molecular Dynamics (MD) studies by \citet{Fredon:2017, Fredon:2018, Fredon:2021A} resulting in an expression for the desorption probability 
\begin{equation}
P_\text{CD}^i= \textit{f}\left(1 - \exp{\left(-\frac{\chi^i \Delta H_\text{react}-\left|E_\text{bind}^i\right|}{3\left|E_\text{bind}^i\right|}\right)}\right)
\label{eq:PcdFredon}
,\end{equation}
where $\Delta H_\text{react}$ is reaction enthalpy, $\chi^i$ is the fraction of the reaction enthalpy that goes into translational excitation of product $i$, $E_\text{bind}^i$ is the binding energy of product $i$ to the surface, and an empirical $\textit{f}$-factor which is recommended to be set at 0.5.  
Direct information on $\chi$ cannot be directly obtained from the classical MD simulations and has to be constrained from either laboratory experiments or quantum chemical simulations \citep{Pantaleone:2020, Ferrero:2023A} that treat one particular system in detail. The MD simulations however show that convertion between rotational, vibrational, and translational excitation is slow, and an equipartition approximation is hence not justified \citep{Fredon:2021B}. Astrochemical models showed that $\chi= 0.1$ is most consistent with astronomical observations of several COMs in the gas phase. 
\citet{Furuya:2022B} constrained $\chi$ to 0.07 based on simulations of  laboratory experiments of the  chemical desorption of HS and \ce{H2S} \citep{Oba:2018, Oba:2019} and \ce{PH2} and \ce{PH3} \citep{Nguyen:2020, Nguyen:2021A}. They compared the different expressions for the photodesorption probabilities and found that the theoretical description of \citet{Fredon:2021B} (Eq.~\ref{eq:PcdFredon}) was most consistent with the experiments.

\section{SURFACE CHEMISTRY}
\label{sec:surfreact}

\subsection{Surface reaction mechanisms}
\label{sec:reactions}
Grain surface chemistry is the result of different -- often competing -- processes. Species that land on a grain can react with other species once they meet. For surface reactions, the transport of species to allow them to meet is usually the rate-limiting step.
In the past, three reaction mechanisms were considered (see \textbf{Figure~\ref{fig:surfmechs}}): the Langmuir-Hinshelwood, the Eley-Rideal, and the hot-atom mechanism. Recently, a non-diffusive mechanism was added as a fourth mechanism. The diffusive Langmuir-Hinshelwood mechanism is important for reactions with H or \ce{H2} which are very mobile on an ice surface.  Here reactants meet through the diffusion of at least one reactant (see Section \ref{sec:diffusion}).
In the Eley-Rideal mechanism, a (stationary) reactant is hit by another species from the gas phase. This is rare for low surface coverage, that is when a surface is not fully covered by molecules and radicals, but can be important, for instance, for surface reactions involving CO when reactants land on a CO-coated ice mantle. The hot-atom mechanism is a combination of both mechanisms. Non-thermalized species travel some distance over the surface to meet other species. This is typically considered important when the gas is warmer than the surface, as is often the case in experiments, and other types of energy sources can trigger this mechanism as well. This can for instance be the adsorption energy of freshly landed species or the excess energy of the reaction in which the species was formed.
The non-diffusive mechanism, finally, is a two-step process where a stable species becomes reactive by a reaction or energetic processing event and subsequently reacts with another reactant in close vicinity. Here no diffusion is required since the reactants are formed close to each other. Examples are hydrogenation or abstraction reactions that create radicals in close vicinity or dissociation reactions by UV irradiation or impacting cosmic rays. Microscopic models like kinetic Monte Carlo models \citep{Cuppen:ChemRev} automatically consider the Langmuir-Hinshelwood, Eley-Rideal and non-diffusive mechanisms, and the final chemical evolution is the result of the competition between the different reactions through these mechanisms. Through these simulations the non-diffusive mechanism was found to be the dominant route to form ethylene glycol and glycolaldehyde starting from the hydrogenation of CO \citep{Fedoseev:2015A, Simons:2020}. For the more-standard rate equation models the different mechanisms have to be added to the equations specifically. The non-diffusive mechanism is then included in a two-step way. \citet{Garrod:2011} did this initially for the reaction of CO + OH where the rate constant of this reaction depends on the formation of OH through, for instance, H + O. A comprehensive way of describing this process can be found in \citet{Jin:2020}.

\begin{figure}
 \includegraphics[width=\textwidth]{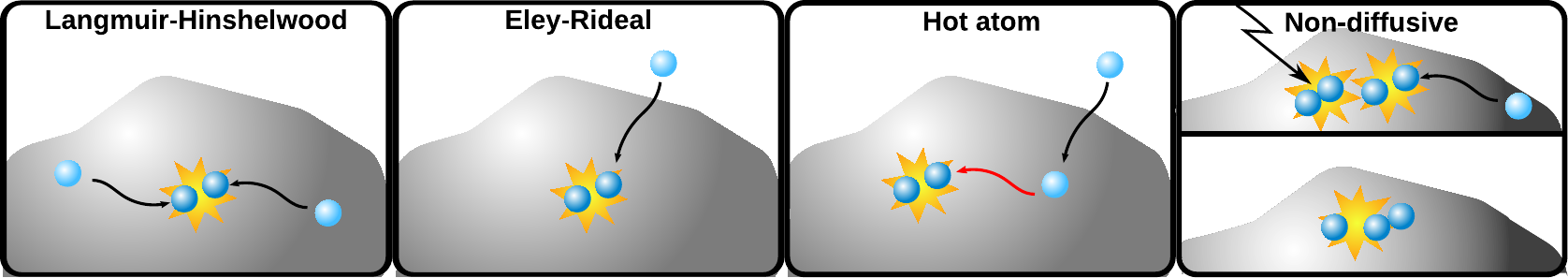}
\caption{Different surface mechanisms for reactants to meet to allow reaction. For Langmuir-Hinshelwood, one or both reactants diffuse over the surface to meet, Eley-Rideal occurs when a species from the gas phase directly hits a reactant on the surface, the hot-atom mechanism combines both earlier mechanisms where the incoming species is more mobile, and the non-diffusive mechanism finally involves two steps, where radicals are in each others vicinity by a previous reaction or a dissociation event after which they can react together without the need for the radicals to move.}
\label{fig:surfmechs}
\end{figure}

Once two reactants meet, surface reactions can occur. At the low temperatures relevant for ice formation only exothermic reactions take place with no or very low reaction barriers. 
Whether a reaction occurs, depends also on other processes such as diffusion and desorption which set the time that two reactants can interact and attempt to react. 
The outcome of a reaction also depends on the orientation in which two reactants meet. A good example is HCO + H. This might lead to the abstraction of H to form \ce{CO + H2} when the hydrogen atom is close to the H atom, to the hydrogenation to \ce{H2CO} when the H atoms arrives on C-atom side, or to no reaction when the H atom approaches on the O side of the molecule. This is reflected in the branching ratio. For reactions without a barrier, these branching ratios depend on the geometrical orientation  \citep{Lamberts:2018}. The branching ratios for reactions with a barrier are determined by a combination of geometrical orientation and the details of the crossing of the barriers of all individual channels. 

Many molecules are formed on grain surfaces through radical-radical reactions where the radicals can be formed from stable species in different ways:
\begin{description}
 \item[hydrogenation reactions] \ce{A + H. -> HA.}
 \item[hydrogen-abstraction reactions] \ce{HA + H. -> A. + H2} or \ce{HA + OH. -> A. + H2O}
 \item[energetic processes] \ce{AB + $h\nu$ -> A. + B.} or \ce{AB + CR -> A. + B.}
\end{description}
The first two reaction types, typically involve a barrier, but --as discussed in Section~\ref{sec:method_comp}-- quantum chemical tunneling can make these barriers less prohibitive. In hydrogenation reactions and hydrogen-abstraction reactions, it is the light H atom that is involved in most of the movement, even if the abstracter is \ce{OH.}. Another possibility to create radicals is by energetic processing such as UV irradiation or by cosmic rays; these provide the energy required to break bonds in the initially stable molecule. Once these radicals are formed, they can react together through radical-radical reactions to increase in molecular complexity. Laboratory experiments proved the formation of methylamine and glycine in water-rich ice \citep{Ioppolo:2021}, glycerol and ethylene glycol in CO-rich ices \citep{Chuang:2016, Chuang:2017}, and series of different COMs in bulk ice layers by energetic processing \citep{Zhu:2020, Oberg:2009A} but this general mechanism. 

Since surface reactions are the result of competition between many processes, it is hard to obtain reaction rate constants or rate coefficients directly from experiments. These rate coefficients are however required for astrochemical models. This is where quantum chemical calculations come into play. Since they can focus on single processes, it is easier to obtain information on a specific reaction directly, without the influence of other processes. Of course, quantum chemical calculations have their own caveats and ideally, a joined theoretical and experimental approach should be taken.

\subsection{Reactions in water-rich ice}
\label{sec:water-rich}
We discuss different surface reactions that follow the evolution of interstellar ice mantles, as shown in \textbf{Figure~\ref{fig:icelayers}}. This subsection  first focuses on water-rich ices and Section~\ref{sec:CO-rich} continues with CO-rich ices. 
\textbf{Figure~\ref{fig:fluxH2Orich}a} shows a network of reactions leading to the formation of \ce{H2O}. The graph is based on a kinetic Monte Carlo simulation of a dark cloud with a density of $2\times 10^4$~cm$^{-3}$. The boxes in the plot indicate the different species, with the colored arrows leading to the box giving its formation reactions and black arrows pointing away from the boxes the destruction reactions. The thickness of the arrows indicates the relative occurrence of the reactions. Only reactions that contribute at least 5\% to the formation or destruction are included. As mentioned in Section~\ref{sec:observations}, the gas phase is mostly atomic at this point except for \ce{H2}. Oxygen atoms (in red) lands on the grain and reacts with H to form OH and with another O to from \ce{O2}. Judging by the width of the arrows the \ce{O + H} reaction occurs more frequently. 
Although the formation routes of different species are treated separately in the following section, \textbf{Figure~\ref{fig:fluxH2Orich}} shows that different types of chemistries are interconnected. OH, for instance, plays a role in the formation of \ce{H2O}, \ce{CO2}, and \ce{HCOOH}.

\begin{figure}
    \centering
    \includegraphics[width=0.8\textwidth]{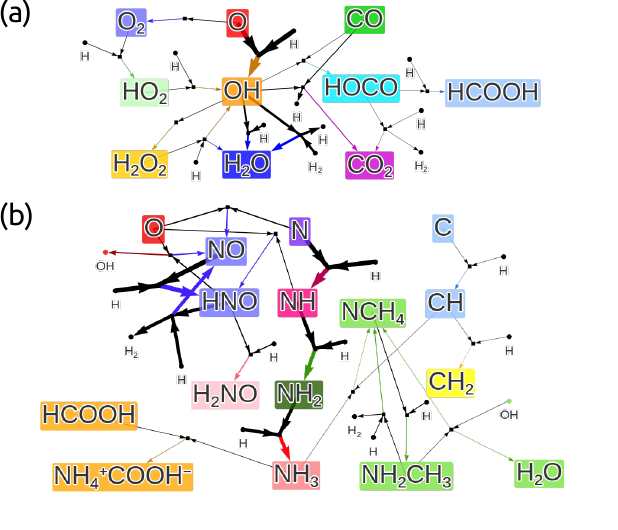}
    \caption{Reaction network relevant for the formation of (a) water and (b) nitrogen-containing species. The squares represent the different reactants and the thicknesses of the arrows approaching and leaving the squares indicate reaction occurrences; only reactions contributing at least 5\% to the formation or destruction are included. Both networks are constructed based on the same kinetic Monte Carlo simulation of a dense cloud with a density of $n_\text{H}= 2\cdot 10^4$~cm$^{-3}$, $T=12$~K, $A_\text{V}=95$~mag, and a collapse time of $10^5$~yrs. The chemical network reported in \citet{Ioppolo:2021} is used.}
    \label{fig:fluxH2Orich}
\end{figure}

\subsubsection{\ce{H2O} formation}
Water ice can form through  hydrogenation of atomic oxygen (O), molecular oxygen (\ce{O2}), and ozone (\ce{O3}) \citep{Tielens:1982}. The relative importance of the different routes depends on O/H  ratio. Only the \ce{O} and \ce{O2} routes appear in \textbf{Figure~\ref{fig:fluxH2Orich}a} due to the low O/H ratio in the simulations. If a particular formation route closes due to physical constraints, another route is taken \citep{Cuppen:2007A, Wakelam:2013}. This shows the importance of studying many different reactions and including all in the grain surface models since their relative importance varies with for example gas-phase composition. Three experimental groups published independently and around the same time work on the \ce{O2} route \citep{Miyauchi:2008,Ioppolo:2008,Matar:2008}. The reaction \ce{H + O2 -> HO2} was shown to be much more efficient than previously included in models and tunneling was found not to play a role in this initial reaction. Calculations of gas-phase rates for atmospheric purposes had indeed shown that the reaction is barrierless when H and \ce{O2} hit each other at the correct angle. The experiments further showed an increasing yield of \ce{H2O2} upon increasing temperature until close to the \ce{O2} desorption temperature \citep{Ioppolo:2010}. This showed that H-atom desorption is not the dominant process at high temperatures, but that H can penetrate into the \ce{O2} ice and react within the bulk of the ice, leading to a very different picture from \textbf{Figure~\ref{fig:surfmechs}}, which only considers reactions occurring on the \emph{surface} of ice grains. The experiments further showed that the three hydrogenation channels are highly interconnected.  \citet{Cuppen:2010B} used co-deposition experiments with different ratios of H and \ce{O2} to probe different stages of the network  and they found indeed extra pathways connecting the three channels, in particular, \ce{HO2 + H -> O + H2O}.

Whereas, the initial reaction to form \ce{H2O2} through \ce{HO2} was near-barrierless, the reactions \ce{H + H2O2 -> H2O + OH} and \ce{H2 + OH -> H2O + H} further down the reaction network have a barrier.  \citet{Oba:2014a} studied the \ce{H + H2O2 -> H2O + OH} reaction and they found a clear kinetic isotope effect (KIE) of roughly 50 in formation of water between hydrogenation and deuteration of \ce{H2O2} as well as for \ce{D2O2}. The computational KIE determined by \citet{Lamberts:2016II} is at least a factor of four higher than the experimental KIE since it is based solely on the rate constant for the reactions, excluding any additional isotopic effects such as a difference in binding, diffusion, and sticking. 
The reaction \ce{H2 + OH -> H2O + H} has eight different permutations using HD, \ce{H2}, \ce{D2}, OH, and OD. Also for this reaction, a KIE is observed, indicating that tunneling is important in this reaction where the rate constant for the formation of \ce{H2O} from OH is 350--400 times higher than for the formation of \ce{HDO} from OH \citep{Oba:2012,Meisner:2017}. For \ce{H + OH} and \ce{D + OH} reactions this effect is not expected, and hence we would hence expect deuterium fractionation to be less efficient in regions with a low H/\ce{H2} ratios, i.e., at higher densities.

\subsubsection{CO$_2$ formation}
After \ce{H2O}, carbon dioxide (\ce{CO2}) is one of the most abundant solid-phase species in the interstellar medium, with abundances between 20 and 30\% of \ce{H2O} ice, depending on the targets observed \citep{Boogert:2015}. In dense molecular clouds, the \ce{CO2} formation is triggered primarily through the \ce{CO + OH} surface reaction \citep{Oba:2010, Ioppolo:2011b, Noble:2011}. \ce{CO2} is found to occur in both \ce{H2O} and \ce{CO}-rich ices \citep{Pontoppidan:2008}, which can be rationalized by realizing that it is formed by reactions between a species linked to CO ice (CO or HCO) and a precursor of water (O or OH). Some of it is hence likely situated at the transition region between the \ce{H2O}-rich phase and the CO-rich layer, which is confirmed by its intermediate threshold $A_\text{V}$.

The \ce{CO + OH} reaction proceeds by the initial formation of the HOCO complex which is highly exothermic. Dissipation of this excess energy leads to a stable long-lived intermediate and the barrier towards \ce{CO2 + H} is too high to direcly form \ce{CO2} \citep{Arasa:2013, Molpeceres:2023}. The HOCO intermediate can, however, react with an additional  hydrogen atom to form \ce{H2 + CO2}, HCOOH, or \ce{H2O + CO} (see \textbf{Figure~\ref{fig:fluxH2Orich}a}) as is confirmed experimentally and computationally \citep{Goumans:2008, Ioppolo:2011a, Ioppolo:2011b, Qasim:2019} and hence \ce{CO2} is formed in a two-step process.  The HOCO complex is also an important intermediate to the formation of glycine under dark conditions \citep{Ioppolo:2021}.

\citet{Chang:2012} studied the surface reaction \ce{CO + O + H} by means of a Unified Microscopic-Macroscopic Monte Carlo simulation of gas-grain chemistry in cold interstellar clouds. In their model, solid \ce{CO2} is produced mainly by the reaction \ce{CO + OH}, which occurs by a so-called ``chain reaction mechanism'', in which an H atom first combines with an O atom lying above a CO molecule, so that the OH does not need to undergo horizontal diffusion to react with CO, which is similar to the non-diffusive mechanism suggested by \citep{Garrod:2011}. This scenario is not far from the experimental conditions described in \citet{Ioppolo:2013a}, where O and H atoms meet to form OH radicals that then further react with neighboring CO molecules to form \ce{CO2}. \cite{Chang:2012} concluded that CO accreted on water-rich dust grains is mainly converted into \ce{CO2} by this reaction, but that \ce{CO2} formation becomes inefficient at later times, leading, for the low-mass protostar case, to a layer of almost pure CO, with some conversion to formaldehyde and methanol. \ce{CO2} formation in a CO-rich ice is discussed in Section \ref{sec:CO-rich}.

\subsubsection{N-bearing molecules formation} 
Although ammonia (\ce{NH3}) is among the most ubiquitous species in the gas phase in space after \ce{H2} and CO, the detection of solid \ce{NH3} has remained elusive and/or controversial for decades  
until it was clearly detected in 24 low-mass YSOs by the \textit{Spitzer c2d} team with abundances of $\sim2$\% to 15\% with respect to \ce{H2O}, and an average abundance of 5.5\% $\pm2.0$\% \citep{Bottinelli:2010}. A combination of observations and laboratory studies suggests that ammonia is  primarily present in \ce{H2O}-rich ices and forms upon hydrogenation of atomic N together with water ice formation in a relatively low-density molecular phase of star formation \citep[see \textbf{Figure~\ref{fig:fluxH2Orich}b}; ][]{Bottinelli:2010}. 

\textbf{Figure~\ref{fig:fluxH2Orich}b} shows the reaction network involving nitrogen surface chemistry. It shows the connectivity of nitrogen chemistry to several other chemistries through its reaction with O and CH, forming hydroxylamine (\ce{NH2OH}) and methylamine (\ce{CH3NH2}), as well as its role in salt formation (e.g., \ce{NH4+HCOO-}). A catalytic cycle around HNO can be observed, where the HNO radical acts as a catalyst for the formation of OH by reaction with atomic O or \ce{H2} by reaction with H atoms \citep{Penteado:2017}. 
The sequential hydrogenation of N atoms was first investigated mass spectrometrically by \citet{Hiraoka:1995} who hydrogenated N atoms trapped in a matrix of solid \ce{N2}. \citet{Hidaka:2011} spectroscopically confirmed the formation of ammonia in a solid \ce{N2} matrix at cryogenic temperatures. In 2015, \citet{Fedoseev:2015B}  studied the surface formation of \ce{NH3} through sequential hydrogenation of N atoms at 15~K. They found the reactions to proceed through a Langmuir-Hinshelwood mechanism that is fast and likely barrierless, thus confirming previous findings. In a follow-up study deuterium enrichment of ammonia ice was found to likely be caused by the higher binding energy of D atoms to the ice surface compared to H atoms \citep{Fedoseev:2015C}. The detection of isocyanic acid (HNCO) in ices containing CO molecules highlighted the reactivity of the intermediate NH and \ce{NH2} species during the formation of \ce{NH3} \citep{Fedoseev:2015B}. 

Hydroxylamine (\ce{NH2OH}) is a potential precursor of complex prebiotic species in space \citep[glycine and $\beta$-alanine; ][]{Blagojevic:2003}. In their combined laboratory and modeling work, \citet{He:2015B} reported the formation of hydroxylamine on an amorphous silicate surface via the oxidation of \ce{NH3}. Since ammonia resides mostly in a water-rich mantle layer, this reaction is likely to occur within such an ice layer. Hydroxylamine can also be formed through the hydrogenation of nitric oxide (NO) via several barrierless reactions, as was confirmed on several different interstellar relevant substrates \citep{Congiu:2012A, Congiu:2012B, Fedoseev:2012}. Up to now, there is only a single detection of \ce{NH2OH} in the gas phase indicating that hydroxylamine is mostly directly converted to other species on the ice surface before desorption or, alternatively, efficiently consumed through gas-phase reactions with \ce{H3^+} and \ce{CH5^+} \citep{Blagojevic:2003, Pulliam:2012, Rivilla:2020}.

\subsubsection{CH$_4$ formation}
Methane (\ce{CH4}) is one of the other few species besides water that have been unambiguously identified in the solid phase in molecular clouds. With abundances of 2\% to 10\% of that of solid water, \ce{CH4} ice is not as abundant as \ce{CO2}. Still, \ce{CH4} is considered an important player in prebiotic chemistry \citep{Kobayashi:2017}. From observational surveys of \ce{CH4} ice towards low- and high-mass young stellar objects, it could be deduced that much of the \ce{CH4} is expected to be formed by the hydrogenation of C on dust grains \citep{Oberg:2008}. Moreover, it appears that \ce{CH4} ice is strongly correlated and mixed with solid \ce{H2O} in interstellar ices \citep{Qasim:2020a}.

The solid-state formation of methane has typically been assumed to follow four sequential atomic hydrogenation steps of the carbon atom in the \ce{^3P} ground state ever since this was already postulated in the late 1940s \citep{Hulst:1949,dHendecourt:1985,Brown:1991}. Recently, this route has been confirmed experimentally \citep{Qasim:2020a,Qasim:2020b} through the simultaneous use of well-characterized C- and H-atom beams: 
\ce{^3C + H -> CH}, \ce{CH + H -> CH2}, \ce{^3CH2 + H -> CH3}, and \ce{CH3 + H -> CH4} (see \textbf{Figure~\ref{fig:fluxH2Orich}b}).
However, it should be noted that the H-atom beam contains a non-negligible amount of \ce{H2}. 
Therefore, surface reactions involving molecular hydrogen need to be considered as well, as was done by \citet{Krasnokutski:2016} and \citet{Henning:2019}.
A recent combined experimental and theoretical work \citep{Lamberts:2022} shows that \ce{CH4} can be formed by combining C atoms with only \ce{H2}/\ce{D2} on amorphous solid water at low temperatures. They concluded that \ce{H2} plays a more important role in the solid-state formation of methane than assumed so far because, although not barrierless for all binding sites on water, the reaction \ce{C + H2 -> CH2} can take place under dense cloud conditions, reaction \ce{CH + H2  -> CH3} is barrierless, and reactions \ce{CH2 + H2 -> CH3 + H} and \ce{CH3 + H2 -> CH4 + H} can take place via a tunneling mechanism. The latter implies that  the deuterium fractionation of methane is more complex than previously assumed. 

\subsubsection{\ce{CH3OH} formation}
\citet{Qasim:2018} demonstrated that the formation pathway of methane is linked to the formation of methanol (\ce{CH3OH}) ice in a water-rich layer by the sequential surface reactions \ce{CH4 + OH -> CH3 + H2O} and \ce{CH3 + OH -> CH3OH}. 
As will be discussed later in the text, \ce{CH3OH} is thought to be mainly formed through surface reactions in a CO-rich ice. The \ce{CO + H} channel is 20 times more efficient at forming \ce{CH3OH} than the \ce{CH4 + OH} channel at temperatures around 10 K. However, the \ce{CH4 + OH} channel can explain the observation of \ce{CH3OH} at low visual extinctions \citep{Boogert:2013}, that is prior to the heavy CO freeze-out stage of dense cold cores, which suggests that COMs can form in both \ce{H2O}- and CO-rich ices.

Another \ce{CH3OH} formation pathway proceeds through the insertion of oxygen atoms in their first electronically excited state [O(\ce{^1D})] into methane to form methanol in astrophysical ice analogs \citep{Bergner:2017}. Gas-phase insertion of excited O(\ce{^1D}) into \ce{CH4} has been shown experimentally to be essentially barrierless \citep{DeMore:1967}, and indeed theoretical studies suggest a small $\sim$280~K barrier \citep{Hickson:2022}. \citet{Bergner:2017} used a deuterium UV lamp filtered by a sapphire window to selectively dissociate \ce{O2} within a mixture of \ce{O2}:\ce{CH4} and observed efficient production of \ce{CH3OH} via O(\ce{^1D}) insertion. They suggested this as a potentially efficient mechanism for the insertion of oxygen in other species to form COMs in cold cores, where the internal cloud UV field can provide enough photons to form O(\ce{^1D}) in ices, as well as in the midplane of protoplanetary disks. A limiting factor for both \ce{CH4} pathways towards \ce{CH3OH} is the availability of \ce{CH4} in the ice. For this reason, these reactions do not appear in \textbf{Figure~\ref{fig:fluxH2Orich}}, since in the O-rich environment of these simulations little \ce{CH4} is present. The reaction is likely more important in more C-rich environments.

\subsubsection{COM formation}
The combined laboratory and modeling work by \citet{Ioppolo:2021} show that glycine can form in a water-rich ice layer from the recombination of radicals and molecules formed in the proximity of each other with only hydrogen diffusing at 10~K in the ice. 
Both a microscopic kinetic Monte Carlo (kMC) model that treats the surface chemistry in detail and can be directly compared to the experiments, as well as a rate equation model that can treat time-dependent physical conditions and can be more easily compared to astronomical observations, were applied. These models show that glycine forms through the non-diffusive reaction of \ce{NH2CH2}, a precursor of methylamine (\ce{NH2CH3}), with HOCO. The production of \ce{NH2CH2} radicals has a substantial contribution from the reaction of CH, a precursor of \ce{CH4}, with ammonia (see \textbf{Figure~\ref{fig:fluxH2Orich}b}). 
Following the same non-diffusive mechanism, in principle, once formed, other functional groups can be added to the glycine backbone by hydrogenation abstraction reactions, resulting in the formation of other amino acids, such as alanine and serine in dark clouds in space \citep{Oba:2015}. 

Although only an upper limit for glycine ice in the ISM is currently available \citep{Gibb:2004}, glycine has been detected together with methylamine, one of its possible precursors, and other organic compounds in the coma of comets in the Solar system
\citep{Elsila:2009, Altwegg:2016}. There are several indications that the material sublimated from 67P/Churyumov-Gerasimenko (67P/C-G) is pristine and has an interstellar origin \citep{Cleeves:2014,Hadraoui:2019}. These conclusions strengthen the importance of the \ce{NH2CH2 + HOCO} formation channel for glycine in space without excluding other energetic routes at later stages of star formation. A prestellar formation of glycine and potentially other amino acids in space shifts the search for life-relevant COMs to much earlier stages of star formation with implications on the degree of complexity that organic material can reach in the ISM \citep{Yang:2022, McClure:2023}. Moreover, the non-diffusive mechanism has been recently implemented in gas-grain models with the result of moving the production of COMs from the warm-up phase of protostellar envelops almost exclusively to also cold dark cores \citep{Jin:2020, Garrod:2022}.

\subsection{Reactions in CO-rich ice}
\label{sec:CO-rich}
When gas densities increase substantially in dense cold cores, depletion of gas-phase material onto the grains becomes rapid. At the center of a high-density collapsing core, the top ice layer consists of predominantly CO ice as the result of ``catastrophic'' CO freeze-out \citep{Pontoppidan:2006, Pontoppidan:2008}. The CO ice layer covers water ice grain mantles and initiates a rich chemistry leading to the formation of many different COMs through hydrogenation addition and abstraction reactions.

\subsubsection{\ce{CO2} formation}
\ce{CO2} cannot only be formed through the subsequent reactions \ce{CO + OH -> HOCO} and \ce{HOCO + H -> CO2 + H}, as mentioned in Section~\ref{sec:water-rich}, \ce{CO2} can also form through other reaction routes such as  \ce{HCO + O} and \ce{CO + O}. In their DFT calculations \citet{Goumans:2008} investigated the formation of \ce{CO2} in the gas phase and on coronene via the three aforementioned pathways. The \ce{HCO + O -> HCO2} reaction was found to be barrierless  and to be more than sufficiently exothermic to subsequently cleave the H--C bond, to form \ce{CO2}. This reaction is not yet experimentally confirmed since it is challenging to investigate in the laboratory.  Other \ce{CO2} surface formation reaction routes will compete and HCO is hard to form at sufficient quantities without reacting further to  formaldehyde and methanol. 

The reaction \ce{CO + O -> CO2} has a substantial barrier of $> 2000$~K in the gas phase \citep{Talbi:2006, Goumans:2008} and because it involves rather heavy species, the onset temperature for tunneling is at a too low temperature for the reaction to significantly contribute to the formation of solid \ce{CO2} under interstellar conditions \citep{Goumans:2010a}. Indeed experiments showed that this reaction yields a low amount of \ce{CO2} \citep{Raut:2011}. Atomic oxygen reacts preferentially with other O atoms to form \ce{O2} and \ce{O3} instead of \ce{CO2}. Under the same experimental conditions \ce{CO2} is found to form more efficiently through the \ce{CO + OH} route. These results are in good agreement with other experimental results \citep{Raut:2011, Oba:2010, Ioppolo:2011b, Noble:2011}, as well as with astrochemical models and observations showing a link between water and \ce{CO2} \citep{Chang:2012}. 

\subsubsection{\ce{CH3OH} formation}
It has been shown that formaldehyde (\ce{H2CO}) and methanol form efficiently at 10 K through the sequential hydrogenation of CO ice at \ce{10-15}~K \citep{Hidaka:2009, Fuchs:2009, Cuppen:2009, Rimola:2014}. Deuteration experiments were also performed on CO ice confirming the formation of both fully deuterated formaldehyde and methanol, although with substantially lower reaction rates \citep{Nagaoka:2005, Watanabe:2006}.
\textbf{Figure~\ref{fig:fluxCOrich}} shows an example network for the reactions involving CO. 
Here again the thickness of the arrows indicates the occurrence of a particular reaction as observed in a kinetic Monte Carlo simulation during $10^4$~yrs of free fall collapse with an initial density of $n_\text{H}= 2\cdot 10^5$~cm$^{-3}$, $T=12$~K, and $A_\text{V}=95$~mag \citep{Holdship:2016}. The input reaction rate constants for all reactions were determined from quantum chemical calculations that explicitly took tunneling into account \citep{Andersson:2011, Song:2017, Alvarez-Barcia:2018}. The figure clearly shows that there is an active cycling between CO, HCO, \ce{H2CO} through hydrogenation and hydrogen abstraction \citep{Chuang:2016}. In fact, the abstraction reactions become the main route to form \ce{H2} under these circumstances. Because of this cycling, the radical concentration is relatively high and radicals are formed at many different locations, albeit quickly destroyed again, which supports the non-diffusive mechanism. Radicals can then be used to form more complex molecules as is discussed in the next section. They can also act as hydrogen abstractors as is the case in the reaction between \ce{CH3O} and \ce{H2CO} to from \ce{CH3OH}. It was recently confirmed that this can indeed be an important alternative route to form \ce{CH3OH} depending on the conditions \citep{Simons:2020, Santos:2022}.

\subsubsection{COM formation}
A series of laboratory and modeling work showed that other COMs, such as glycolaldehyde (\ce{CH2OHCHO}), ethylene glycol (\ce{HOCH2CH2OH}), glyceraldehyde (\ce{HOCH2CH(OH)CHO}), glycerol (\ce{HOCH2CH(OH)CH2OH}), methyl formate (\ce{HC(O)OCH3}), and methoxymethanol (\ce{CH3OCH2OH}), are all formed through non-diffusive radical-radical recombination reactions at 10~K \citep{Woods:2013, Fedoseev:2015A, Chuang:2016, Chuang:2017, Fedoseev:2017, He:2022A}. Those papers indicate a `non-energetic' way to form the precursors of ribose, a simple sugar, and xylitol, a sugar alcohol, in dense molecular clouds and, at the same time, some of the methyl formate isomers that are abundantly observed in the gas phase in the ISM \citep{El-Abd:2019}. Moreover, experiments show that the addition of acetylene (\ce{C2H2}) to the \ce{CO + H} reaction network can lead to the formation of propanal (\ce{CH3CH2CHO}) and 1-propanol (\ce{CH3CH2CH2OH})  --three carbon-bearing representatives of the primary linear aldehydes and alcohols, respectively-- under cold dark conditions \citep{Qasim:2019}. Computational work starting from the CCH radical studied a pathway to ethanol (\ce{CH3CH2OH}) \citep{Perrero:2022B}.
The hydrogenation of accreting C atoms and CO molecules on ice grains results in ketene (\ce{CH2CO}), acetaldehyde (\ce{CH3CHO}), and possibly ethanol  under dark cloud conditions \citep{Fedoseev:2022, Perrero:2023, Ferrero:2023B}.

\begin{figure}
    \centering
    \includegraphics[width=0.8\textwidth]{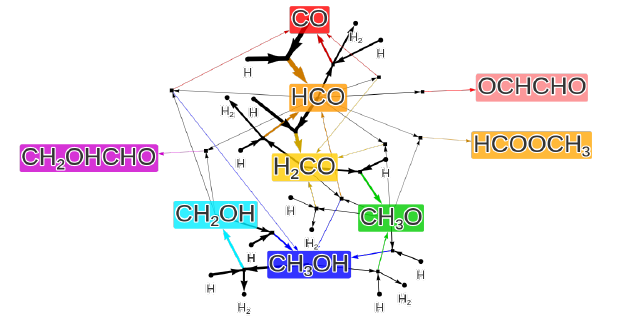}
    \caption{Reaction network on the formation of methanol (\ce{CH3OH}). The thickness of the arrows indicates the occurrence of the different reactions; reactions that contribute less than 5~\% to the formation or destruction are left out. The network is constructed based on a kinetic Monte Carlo simulation of the freeze-out phase after a cloud with a density of $n_\text{H}= 2\cdot 10^5$~cm$^{-3}$, $T=12$~K, $A_\text{V}=95$~mag, has collapsed over $10^4$~yrs.}
    \label{fig:fluxCOrich}
\end{figure}

\subsection{Thermally activated reactions and salt formation}
\label{sec:bulk}
Initially, most ice chemistry occurs on the surface of the ice mantle or interstellar grain. Species land from the gas phase on the surface, diffusion occurs more easily at the surface, and even in the case of non-diffusive chemistry the initial reaction that creates one or both radicals often involves a hydrogenation or abstraction reaction that typically occurs at the surface. However, once a thick ice mantle is formed and most of the heavy species in the gas phase are depleted, bulk processes become more important, especially if the temperature of the ice mantle increases. There is a class of reactions that is not limited by the availability of new radical species through adsorption, reaction, or dissociation but involves common ice species such as \ce{NH3}, \ce{HCOOH}, or even \ce{H2O}, which are abundantly present. Their reaction rate constants are small because of the reaction barrier that needs to be crossed. However, these reactions become competitive with the standard surface reactions in the case of thick ice layers at slightly elevated temperatures, hence the name ``thermal reactions''.

\citet{Theule:2013} give an overview of the different classes of thermally activated reactions and here we extend on this work. One of these classes is acid-base reactions leading to salts. Acids are good \ce{H+} donors and bases good \ce{H+} acceptors. Most of the thermal reactions are relevant to the warm-up phase (see \textbf{Figure~\ref{fig:icelayers}}), but some of the acid-base reactions have such low activation barrier that they can already proceed at low temperatures, for instance during the core collapse phase.  An example are acid-base reactions such as \ce{NH3 + HCOOH -> NH4+HCOO-}, which has a low activation barrier of 70~K \citep{Theule:2013,Bergner:2016} (see \textbf{Figure~\ref{fig:fluxH2Orich}b}). When the resulting salt desorbs, it is released in the gas phase as neutrals \ce{NH3} and HCOOH \citep{Kruczkiewicz:2021}. 
Among the common ice species, there are only two bases: the strong base \ce{NH3} and the weak base \ce{H2O}. Another possible base is \ce{CH3NH2} but this is not as abundant. For this reason, ammonia salts resulting from reactions with \ce{NH3} will be the most common salts.
\ce{NH4+} cannot be formed and survive in isolation and should always be considered in conjunction with an anion formed from an acid. Possible counterions are \ce{HCOO-}, \ce{CN-}, \ce{OCN-}, and \ce{OH-} formed from HCOOH, HCN, HNCO, and \ce{H2O}, respectively. The latter is amphoteric and can act as both a weak acid and a weak base. The anion \ce{OCN-} has indeed been observed as ice. \ce{NH4+} is observed in ice spectra, even in cold dark molecular clouds \citep{McClure:2023} but given the low activation barrier for the reaction between \ce{NH3} and HCOOH \citep{Theule:2013, Bergner:2016} or HNCO \citep{vanBroekhuizen:2005, Mispelaer:2012, Ratajczak:2009} this is not unexpected. It implies that the concentrations of \ce{NH3} and the reaction acids are high enough that transport is not rate-limiting. The non-detection of HNCO and the observation of \ce{OCN^-} in dense cold cores confirmed by JWST suggest that most of the CO present in a water-rich ice is efficiently converted to \ce{CO2}, a competing channel to the formation of HNCO, and that \ce{OCN^-} can also form through other surface reaction routes in interstellar ice grains \citep{Gibb:2004, vanBroekhuizen:2005, Oberg:2011, Boogert:2022, McClure:2023}. High resolution specroscopic studies of salts embedded in an ice environment deserve more future attention.

Another class of reactions is nucleophilic-electrophile reactions which typically proceed through a series of equilibrium reactions, involving acid-base reactions. Examples are
\begin{align}
\ce{OH- + D2O <=> HDO + OD-}\\
\ce{OD- + H2O <=> HDO + OH-}
\end{align}
which gives as net reaction \ce{H2O + D2O -> 2HDO} but is mediated by  \ce{OH-} ions in the ice \citep{Lamberts:2015}. The barrier for this series of reactions is significantly higher than the salt reactions with 3840~$\pm$~125~K and is hence relevant at higher temperatures between 90-140~K, closed to their desorption limit.
Another example is 
\begin{align}
 \ce{NH3 + CO2} & \ce{<=> NH3+COO-}\\
 \ce{NH3 + NH3+COO- } & \ce{<=> NH4+NH2COO-}\\
 \ce{NH4+NH2COO-} & \ce{<=> NH3 + NH2COOH}
\end{align}
leading to the formation of carbamic acid in a proton-rich environment, \emph{i.e.} a \ce{NH3}- or \ce{H2O}-dominated ice, where the reaction barrier is lowered with respect to a  \ce{CO2}-ice. However, experimentally only mixtures with a \ce{NH3}:\ce{H2O} ratio greater than 1 were considered. In a more dilute environment, which is more representative of interstellar ices these processes become either bulk-diffusion limited or the reaction yield is limited to the cases where the reactants happen to be in close proximity.

Finally, condensation reactions have been observed in interstellar ice analogs. In condensation reactions two molecules form one larger molecule, usually upon loss of a small molecule like \ce{H2O}; hence the name ``condensation''. The peptide bond between two amino acids is formed through a condensation reaction. In interstellar ices, this type of reaction is found to initialize polymerization reactions leading to polyoxymethylene based on \ce{H2CO} \citep{Schutte:1993, Noble:2012b} or polymethylimine using \ce{CH2NH} \citep{Danger:2011}, for instance.

\subsection{Energetic processing: Cosmic rays, UV photons, and free electrons}
\label{sec:energeticproc}
In space, several interstellar radiation sources can interact with ice material. These sources can be the ISRF at the edge of a cloud; secondary UV radiation inside a molecular cloud induced by the interaction of cosmic rays and \ce{H2} gas; the protostar black-body radiation field affecting the nearby gas, ice, and grain material; and the ubiquitous cosmic rays in the ISM \citep[e.g.,][]{Cecchi-Pestellini:1992,  Mathis:1983, Padovani:2018}.
Laboratory experiments can simulate different interstellar radiation types by implementing UV photon sources as described in Section~\ref{sec:photodesorption}, electron sources (e.g., $0.5-5$~keV electron guns), and ion sources (e.g., van de Graaff accelerators, tandem accelerators, and cyclotrons) in UHV setups similar to the one depicted in \textbf{Figure~\ref{fig:UHV_setup}} \citep[see][]{Oberg:2009A, Bertin:2023, Gobi:2017, Urso:2022, Mifsud:2023}. Current laboratory and modeling work show that the formation of amino acids, such as glycine, alanine, and serine, and other prebiotic species can occur by means of energetic (UV photon, cosmic ray, electron, X-ray, and thermal) processing of interstellar relevant ices \citep[e.g.,][]{Bernstein:2002A, MunozCaro:2002}. However, experimental work also shows that the energetic processes that induce the formation of amino acids in space cause their chemical alteration and destruction at higher irradiation doses \citep[e.g.,][]{Gerakines:2015, Mate:2015}. Therefore, the level of chemical complexity reached in interstellar ices will depend on the local interstellar conditions. 

Irradiation of interstellar ices can cause photodesorption as discussed in Section~\ref{sec:photodesorption}, photochemistry, and radiation chemistry. The term photochemistry (also known as photolysis) is typically reserved for processes involving electronic excitation leading to the breaking and reformation of bonds, while radiation chemistry (also named radiolysis) for processes that include both electronic excitation and ionization. Radiolysis requires ionizing radiation with sufficient energy to ionize atoms or molecules in an ice layer by detaching electrons. Ionizing radiation consists of subatomic particles (e.g., $\alpha$ and $\beta$ particles, CRs) or electromagnetic waves (e.g., $\gamma$ photons, X-rays, EUV and VUV photons).  Whereas, the lower energy ultraviolet, and visible light are types of non-ionizing radiation that can only cause excitation, i.e., the promotion of an electron to a higher energy state, in an ice mantle. Unfortunately, the boundary between ionizing and non-ionizing radiation in the ultraviolet spectral region cannot be sharply defined, as different molecules and atoms ionize at different energies between 10 and 33~eV. 

 Experiments simulating energetic processing of interstellar ices to form new molecules via photolysis and radiolysis are routinely performed by many different groups worldwide at both small and large-scale research facilities \citep[e.g.,][]{Herczku:2021, Ciaravella:2020, Paardekooper:2016, Modica:2012, Oberg:2009A}. Decades' worth of laboratory work have proven that both ice photolysis and radiolysis are plausible pathways to chemical complexity in space. Very much in line with the processes described earlier that are triggered upon atom additions, both photolysis and radiolysis form radicals within the ice that can react with other species in the ice. An example of this is shown for methanol ice by comparing Figure 22 of \citet{Oberg:2009A} and Figure 4 in \citet{Zhu:2020}. Although the chemical formation pathways may depend on the type of energetic processing used to irradiate, in this case, UV photons and 5~keV electrons, the same intermediate radicals and several final products are formed in both experimental sets. This highlights that product compositions in photolysis and radiolysis ice experiments are often remarkably similar. Therefore, thus far, there is only evidence for the overall chemical evolution of interstellar ices to be mainly depend on the amount of energy deposited into the ice and not on how it is delivered \citep{Gerakines:2004, Islam:2014, Mate:2015, Mullikin:2021}.
 
 Laboratory processing of methanol ices is a key step to the formation of many complex organic molecules, such as methyl formate (\ce{HCOOCH3}), dimethyl ether (\ce{CH3OCH3}), glycolaldehyde (\ce{HOCH2CHO}), acetic acid (\ce{CH3COOH}), and methoxymethanol (\ce{CH3OCH2OH}) \citep{Bennett:2007A, Bennett:2007B, Palumbo:1999, Modica:2010, Mason:2014, Jheeta:2013, Moore:1996}. Radiolysis studies of methanol ice have been conducted using both high-energy particles (\ce{H+}, \ce{He+} ions, and electrons) as well as low-energy ($<$20~eV) electrons and (6--13~eV) UV photons. In a study comparing methanol ice processing by high-energy (1~keV) electrons with that by low-energy ($<$20~eV) electrons, again the same products were detected \citep{Boyer:2016, Sullivan:2016}. This finding supports the widely accepted idea that high-energy solid-phase radiolysis is mediated by low-energy electron-initiated reactions as a cascade of secondary low-energy electrons is formed along the path of each high-energy particle. In their review on photolysis versus radiolysis of interstellar ices, \citet{Arumainayagam:2019} highlighted the importance of low-energy ($<$20~eV) secondary electrons. The authors further argue that because the ionization threshold is lower in the solid phase than in the gas phase, most photolysis studies of astrochemistry likely involve radiation chemistry. 
 It should be mentioned that, contrary to radiolysis, photolysis triggers, at maximum, a single event per photon, where the energy of the photon is adsorbed by a single species \citep[][and references therein]{Ziegler:2010}. Moreover, the ice thickness affected by different energetic processing largely varies. For instance, UV photons penetrate roughly 100~MLs of \ce{H2O} ice ($\sim$30~nm), 2~keV electrons less than 200 nm, and proton ions with energies between 200-1000~keV implant into 3-30~$\mu$m thick ices \citep{Drouin:2007, Ziegler:2010}. Although much work has been done to investigate pathways of photolysis and radiolysis of solid methanol and other COMs, interstellar molecular tracers that can be used to detect electron/photon-dominated chemistry in the ISM are yet to be identified. 

In their review, \citet{Oberg:2016} compare systematically UV photolysis to X-ray, electrons, and ion radiolysis of interstellar ice analogs and conclude that, although photochemistry is a potential source of prebiotic amino acids and sugars and maybe the original source of enantiomeric excess on the nascent Earth, little is still known on ice photochemistry kinetics and mechanisms. This is also true for radiation chemistry as existing experiments and models have demonstrated the difficulty in extrapolating such kinetics from gas to ice, and between different ice systems \citep[e.g.,][]{Shingledecker:2018B, Shingledecker:2020, Paulive:2021}. An additional complication is that most of the reactions occur in the bulk of the ice instead of on the surface and transport inside the ice is poorly understood. Experiments have so far provided some constraints on energetic processes for specific ice constituents and ice matrices \citep{Materese:2015, Pilling:2010, Jones:2011, Ciaravella:2010}. A more detailed understanding of ice photochemistry and radiation chemistry kinetics is required to predict the typical concentration of COMs in interstellar ices, and thus the abundance delivered to comets and further to nascent planets.

\section{SUMMARY AND FUTURE}
Solid-state laboratory astrophysics has made a strong development over the past two decades. 
Traditionally solid-state laboratory astrophysics was very much geared towards cold cloud conditions using either mass spectrometry or infrared spectroscopy as analysis methods. In recent years, there has been a clear trend away from this in different areas. More different analysis methods are being used, with for instance microscopy techniques like electron microscopy or low-temperature scanning tunneling microscopy.
Photoprocesses are studied over a wider range of wavelengths. This was traditionally mostly in the  UV, but now also includes studies in X-ray and IR regimes. Moreover, lasers are more commonly used which have a much more narrow bandwidth and aid in the molecular understanding of the results.
 An increasing number of laboratory experiments are carried out at international large-scale facilities, such as synchrotrons, neutron scattering, ion accelerators, and free-electron lasers, to access experimental tools and physicochemical conditions not readily available elsewhere. A larger range of different physical conditions is probed and the focus is not solely on cold conditions but also covers disks, comets, and exoplanets. 

The technological advances in both laboratory technology and computational chemistry methods have led to a more molecular understanding of the processes occurring in ice mantles. This has helped in the interpretation of experimental results, drawing similarities between different types of chemistry as well as in the implementation of ice chemistry and physics into astrochemical models.

\begin{summary}[SUMMARY POINTS]
\begin{enumerate}
\item All different ice processes occur in all different environments but their relative importance changes. Initially, surface processes dominate until the gas phase is depleted and the surface area has been reduced due to grain coagulation. Further processing has to occur through bulk chemistry processes, which are triggered by heat, UV, and/or cosmic rays.

    \item From all experimental ice parameters, binding energies are the most complete data set for both stable species and radical species. It is important to use the correct prefactor for desorption.  There is competing evidence whether a universal value between the diffusion barrier and the binding energy exists.

    \item Cold and dark surface chemistry can lead to complex organic molecule formation. This proceeds through radical-radical reactions, where the radicals are formed by dissociation or reactions involving hydrogen. Non-thermal desorption mechanisms like reactive desorption and photodesorption at different wavelengths can return these COMs to the gas phase.

    \item Although radiolysis and photolysis are distinct processes, the final products of their interaction with ices are very similar. A better understanding of ice photochemistry and radiation chemistry kinetics is required to predict the typical concentration of COMs in interstellar ices, and thus the abundance delivered to comets and further to nascent planets. 
    
            \item Development in computational techniques  have made computations on larger ice systems more feasible. These computations give a molecular picture which has greatly enhanced the understanding of the underlying processes and the interpretation of experimental results. 

\end{enumerate}
\end{summary}

\begin{issues}[FUTURE ISSUES]
\begin{enumerate}
    
    \item The increasing number of larger complex organic molecules detected in the ISM  will motivate laboratory astrochemistry to implement a wider range of new experimental techniques to investigate the surface formation and destruction pathways of such molecules, circumventing current experimental limitations in the preparation of such ices and their physicochemical characterization.
    
    \item There are developments in the use of artificial intelligence. This can help in predicting missing experimental data based on molecular features, or create machine learning potentials to speed up expensive DFT calculations.

\item The role of the ice structure and the interaction with the underlying surface in solid-state astrochemistry is still rather unexplored. Ice structure is mostly studied on pure ice surfaces, but the presence of different ice constituents will affect binding sites for instance. The structure of the ice is likely affected by different energy inputs and finally, the possibility of spontelectric effects will impact the electrostatics during reaction. 

\item Experiments and chemical computations show the complexity and details that are involved in ice processes. For the modeling community, it will be a challenge to abstract from this and choose the right level of detail required to improve their models to explain new observations made by, e.g., JWST and ALMA.

\end{enumerate}
\end{issues}

\section*{DISCLOSURE STATEMENT}
The authors are not aware of any affiliations, memberships, funding, or financial holdings that
might be perceived as affecting the objectivity of this review. 

\section*{ACKNOWLEDGMENTS}

We would to thank Joseph Salaris for making the network figures and providing the data required for creating them, Marina Rachid for the figure on the ice abundances, and Will Rocha for the spectroscopy figure. We like to thank the JOYS+ team for the use of the spectrum and fit in Figure~\ref{fig:IR_spectroscopy}.
 We are grateful to Catherine Walsh, Thanja Lamberts, Ewine van Dishoeck, Rob Garrod, Silvia
Spezzano, Naoki Watanabe, Melissa McClure and Albert Rimola for input and discussions in preparation for this review.

%
\bibliographystyle{ar-style2}
\bibliography{refs}

\end{document}

%% file: table_PD.tex
\begin{table}[h]
\tabcolsep7.5pt
\caption{Experimental VUV and IR photodesorption yields of the intact molecule for pure and astronomically relevant ice species.$^{\rm a}$}\label{tab:PD-rates}
\begin{center}
\begin{tabular}{@{}l|c|c|c|c@{}}
\hline
Ice         & Temperature          & Wavelength /          & Photodesorption rate                       & Ref. \\
species     & range                & Irradiation source    &                                            & \\
            & (K)                  & $^{\rm b}$            & (molecules/photon{)}                       & \\
\hline

\ce{H2O}    & $35-100$             & VUV / MWHD            & $(3-8)\times 10^{-3}$                      & {$\rm ^{f}$} \\
            & $18-100$             & VUV / MWHD            & $(1.3\pm0.4 - 4.5\pm1.2)\times 10^{-3}$    & {$\rm ^{g}$} \\
            & $8-90$               & VUV / MWHD            & $(1.3\pm0.2 - 2.5\pm0.7)\times 10^{-3}$    & {$\rm ^{h}$} \\
\ce{D2O}    & $8-90$               & VUV / MWHD            & $(0.7\pm0.1 - 1.5\pm0.3)\times 10^{-3}$    & {$\rm ^{h}$} \\

\ce{CO}     & $15-27$              & VUV / MWHD            & $(2.7\pm1.3) \times 10^{-3}$               & {$\rm ^{i}$} \\
            & $7-15$               & VUV / MWHD            & $(3.5\pm0.5 - 6.4\pm0.5)\times 10^{-2}$    & {$\rm ^{j}$} \\
            & $14$                 & VUV / MWHD            & $(6.4\pm0.2 - 21.2\pm0.3)\times 10^{-2}$   & {$\rm ^{k}$} \\
            & $18$                 & VUV / SYNC            & $(4.1-16)\times 10^{-3}$                   & {$\rm ^{l}$} \\
            & $20$                 & VUV / MWHD            & $(1.4\pm0.7)\times 10^{-3}$                & {$\rm ^{m}$} \\
            & $20$$\rm ^d$         & IR  / FEL             & $(1.1\pm0.3)\times 10^{-8}$                & {$\rm ^{n}$} \\
	    & $18$$\rm ^d$         & IR  / FEL             & $(3-5)\times 10^{-3}$ / 5 and 12 $\mu$m    & {$\rm ^{h}$} \\

\ce{CO2}   & $16-60$               & VUV / MWHD            & $(2.3\pm1.4)\times 10^{-3}$                & {$\rm ^{o}$} \\
           & $10-40$               & VUV / SYNC            & $(1-3)\times 10^{-3}$                      & {$\rm ^{p}$} \\
           & $16-60$               & VUV / MWHD            & $(2.4\pm0.2-2.6\pm0.2) \times 10^{-2}$     & {$\rm ^{q}$} \\

\ce{NH3}   & $8$                   & VUV / MWHD            & $(1.1-4.2)\times 10^{-3}$                  & {$\rm ^{r}$} \\

\ce{CH4}   & $10$                  & VUV / SYNC            & $(2.0\pm1.0) \times 10^{-3}$               & {$\rm ^{s}$} \\
           & $8$                   & VUV / MWHD            & $< 1.7 \times 10^{-4}$                     & {$\rm ^{t}$} \\

\ce{CH3OH} & $20-70$               & VUV / MWHD            & $(2.1\pm1.0)\times 10^{-3}$                & {$\rm ^{u}$} \\
           & $8-130$               & VUV / MWHD            & $< 3 \times 10^{-5}$                       & {$\rm ^{v}$} \\
           & $10$                  & VUV / SYNC            & $< 10^{-5}$                                & {$\rm ^{w}$} \\
	   & $20$                  & IR  / FEL             & $(3\pm1)\times 10^{-8}$                    & {$\rm ^{x}$} \\

\ce{N2}    & $16$                  & VUV / MWHD            & $\sim 2 \times 10^{-4}$                    & {$\rm ^{o}$} \\
           & $14$$\rm ^e$          & VUV / SYNC            & $(1.5-5.3)\times 10^{-3}$                  & {$\rm ^{y}$} \\

\ce{O2}    & $14-21$               & VUV / MWHD            & $(6\pm2)\times 10^{-4}$                    & {$\rm ^{z}$} \\
           & $14$                  & VUV / SYNC            & $(2.1-3.3) \times 10^{-3}$                 & {$\rm ^{y}$} \\

\ce{O3}    & $14$                  & VUV / MWHD            & ($3\pm1)\times 10^{-4}$                    & {$\rm ^{z}$} \\
           & $52$                  & VUV / MWHD            & ($5\pm2)\times 10^{-4}$                    & {$\rm ^{z}$} \\
					
\ce{NO}    & $10$                  & VUV / SYNC            & $(1.1\pm0.4-1.3\pm0.5)\times 10^{-2}$      & {$\rm ^{A}$} \\
         
\ce{H2CO}  & $10$                  & VUV / SYNC            & $(4-10) \times 10^{-4}$                    & {$\rm ^{B}$} \\

\ce{CH3CN} & $15$                  & VUV / SYNC            & $2.5\pm \times 10^{-5}$                    & {$\rm ^{C}$} \\
           & $20$                  & VUV / MWHD            & $< 7.4 \times 10^{-4}$                     & {$\rm ^{D}$} \\
										
HCOOH      & $15$                  & VUV / SYNC            & $< 10^{-5}$                                & {$\rm ^{E}$} \\

\hline
\end{tabular}
\end{center}
\begin{tabnote}
{$\rm ^{a}$} Table extended from \citet{Cuppen:2017}. For some identified ice species, photodesorption rates are stil missing, for other species not identified yet, but likely present as ice is space, photodesorption rates have been already derived. 
{$\rm ^{b}$} Abbreviations: MWHD = Microwave hydrogen-flow discharge lamp {roughly 120-170 nm, including Ly-$\alpha$); FEL = Free Electron Laser (mono-chromatic and on-resonance for selected vibrational mode); SYNC = Synchrotron (monochromatic, tunable in VUV, typically 7-13.6 eV / 91-177 nm, allows to derive photodesorption rates for astronomical regions with different spectral energy distributions). 
{$\rm ^{c}$} For a graphical overview, see Fig. 5 in \citet{Paardekooper:2016} 
($\rm ^{d}$} The difference between these two values is likely due to the use of different ice mixtures: CO:\ce{CH3OH} and CO:\ce{H2O}. 
{$\rm ^{e}$} for CO:\ce{N2} mixed ices. 
{$\rm ^{f}$} \citet{Westley:1995a} 
{$\rm ^{g}$} \citet{Oberg:photoII} 
{$\rm ^{h}$} \citet{Cruz-Diaz:2017} 
{$\rm ^{i}$} \citet{Oberg:photoI, Oberg:2007B}  
{$\rm ^{j}$} \citet{MunozCaro:2010} 
{$\rm ^{k}$} \citet{Chen:2014I} 
{$\rm ^{l}$} \citet{Fayolle:2011B, Bertin:2012}
{$\rm ^{m}$} \citet{Paardekooper:2016} 
{$\rm ^{n}$} \citet{Santos:2023A} 
{$\rm ^{o}$} \citet{Oberg:photoI} 
{$\rm ^{p}$} \citet{Fillion:2014} 
{$\rm ^{q}$} \citet{Sie:2019}
{$\rm ^{r}$} \citet{Martin-Domenech:2017} 
{$\rm ^{s}$} \citet{Dupuy:2017A} 
{$\rm ^{t}$} \citet{Carrascosa:2020} 
{$\rm ^{u}$} \citet{Oberg:2009C} 
{$\rm ^{v}$} \citet{Cruz-Diaz:2016} 
{$\rm ^{w}$} \citet{Bertin:2016} 
{$\rm ^{x}$} \citet{Santos:2023A} 
{$\rm ^{y}$} \citet{Fayolle:2013, Bertin:2013}  
{$\rm ^{z}$} \citet{Zhen:2014} 
{$\rm ^{A}$} \citet{Dupuy:2017B} 
{$\rm ^{B}$} \citet{Feraud:2019} 
{$\rm ^{C}$} \citet{Basalgete:2021III}    
{$\rm ^{D}$} \citet{Bulak:2021} 
{$\rm ^{E}$} \citet{Bertin:2023} 
\end{tabnote}
\end{table}